\documentclass[aps,prd,nofootinbib,twocolumn,superscriptaddress,preprintnumbers,balancelastpage,longbibliography]{revtex4-1}

\usepackage[utf8]{inputenc}
\usepackage{amsmath,amssymb,mathtools,bm}
\usepackage{graphicx, color, hepunits}
\usepackage[dvipsnames]{xcolor}
\usepackage{float}
\usepackage{multirow}
 \usepackage{hyperref} 
\hypersetup{
    colorlinks=true,       
    linkcolor=blue,        
    citecolor=blue,        
    filecolor=magenta,     
    urlcolor=blue          
}
\usepackage[utf8]{inputenc}
\usepackage[english]{babel}
\usepackage{lipsum}

\usepackage{tensor}

\newcommand{\gapp}{g_{app}}

\newcommand{\gann}{g_{ann}}

\newcommand{\DeltaM}{\Delta M}

\newcommand{\es}[2] {\begin{equation} \label{#1} \begin{split} #2 \end{split} \end{equation}}

\begin{document}

\title{Upper Limit on the QCD Axion Mass from Isolated Neutron Star Cooling}

\author{Malte Buschmann}
\affiliation{Department of Physics, Princeton University, Princeton, NJ 08544, U.S.A.}

\author{Christopher Dessert}
\affiliation{Leinweber Center for Theoretical Physics, Department of Physics, University of Michigan, Ann Arbor, MI 48109 U.S.A.}
\affiliation{Berkeley Center for Theoretical Physics, University of California, Berkeley, CA 94720, U.S.A.}
\affiliation{Theoretical Physics Group, Lawrence Berkeley National Laboratory, Berkeley, CA 94720, U.S.A.}

\author{Joshua W. Foster}
\affiliation{Center for Theoretical Physics, Massachusetts Institute of Technology, Cambridge, Massachusetts 02139, U.S.A}

\author{Andrew J. Long}
\affiliation{Department of Physics and Astronomy, Rice University, Houston, TX 77005, U.S.A.}

\author{Benjamin R. Safdi}
\affiliation{Berkeley Center for Theoretical Physics, University of California, Berkeley, CA 94720, U.S.A.}
\affiliation{Theoretical Physics Group, Lawrence Berkeley National Laboratory, Berkeley, CA 94720, U.S.A.}

\date{\today}

\begin{abstract}
  The quantum chromodynamics (QCD) axion may modify the cooling rates of neutron stars (NSs).  The axions are produced within the NS cores from nucleon bremsstrahlung and, when the nucleons are in superfluid states, Cooper pair breaking and formation processes.  We show that four of the nearby isolated Magnificent Seven NSs along with PSR J0659 are prime candidates for axion cooling studies  because {they are coeval, with ages of a few hundred thousand years}
  known from kinematic considerations, and they have well-measured surface luminosities.
  We compare these data to dedicated NS cooling simulations incorporating axions,
  profiling over uncertainties related to the equation of state, NS masses, surface compositions, and superfluidity.  Our calculations of the axion and neutrino emissivities include high-density suppression factors that also affect SN 1987A and previous NS cooling limits on axions. 
  We find no evidence for axions in the isolated NS data, and within the context of the KSVZ QCD axion model 
  we constrain $m_a \lesssim 16$ meV at 95\% confidence.  
  An improved understanding of NS cooling and nucleon superfluidity  could further improve these limits or lead to the discovery of the axion at weaker couplings. 

\end{abstract}
\maketitle

The quantum chromodynamics (QCD) axion is a well-motivated beyond-the-Standard-Model particle candidate that may explain the absence of the neutron electric dipole moment~\cite{Peccei:1977hh,Peccei:1977ur,Weinberg:1977ma,Wilczek:1977pj} and the dark matter (DM) in our Universe~\cite{Preskill:1982cy,Abbott:1982af,Dine:1982ah}.  However, the axion remains remarkably unconstrained experimentally and observationally, despite nearly 45 years of effort dedicated to axion searches (see~\cite{Sikivie:2020zpn} for a review).  The QCD axion is primarily characterized by its decay constant $f_a$, which sets both its mass~\cite{diCortona:2015ldu} \mbox{$m_a \approx 5.7\, \, \mu{\rm eV} \, \big( 10^{12} \, {\rm GeV} / f_a \big)$} and its interaction strengths with matter.
Requiring $f_a$ below the Planck scale implies $m_a \gtrsim 10^{-12} \ \mathrm{eV}$.
The axion mass is currently bounded from above by supernova (SN) and stellar cooling constraints at the level of tens of meV, subject to model dependence and astrophysical uncertainties that are discussed further below.  This work aims to improve upon these upper bounds 
by studying the cooling of old neutron stars (NSs) with ages $\sim$$10^5$--$10^6$ yrs.  

The NS constraints presented in this work are part of a broader effort to probe the QCD axion over its full possible mass range.
Black hole superradiance disfavors QCD axion masses $m_a < 2 \times 10^{-11}$ eV~\cite{Arvanitaki:2009fg,Arvanitaki:2010sy,Cardoso:2018tly}, 
while the ADMX experiment has reached sensitivity to Dine-Fischler-Srednicki-Zhitnitsky (DFSZ)~\cite{Dine:1981rt,Zhitnitsky:1980tq} QCD axion DM over the narrow mass range \mbox{$m_a \sim 2.66$--$3.31$ $\mu$eV} by using the axion-photon coupling~\cite{ADMX:2018gho,ADMX:2019uok}.  Apart from these constraints, and additional narrow-band constraints from the ADMX~\cite{ADMX:2001dbg} and HAYSTAC~\cite{HAYSTAC:2020kwv} experiments at the level of the more strongly-coupled Kim-Shifman-Vainshtein-Zakharov (KSVZ)~\cite{Kim:1979if,Shifman:1979if} axion, there is nearly a decade of orders of magnitude of parameter space open for the axion mass that is un-probed at present.  On the other hand, near-term plans exist to cover experimentally most of the remaining parameter space for QCD axion DM, including ABRACADABRA~\cite{Kahn:2016aff,Ouellet:2018beu,Salemi:2021gck}, DM-Radio~\cite{Silva-Feaver:2016qhh}, and CASPEr~\cite{Budker:2013hfa,Garcon:2017ixh,Aybas:2021nvn} at axion masses $m_a \ll \mu$eV, ADMX and HAYSTAC at axion masses \mbox{$m_a \sim 1$--$100$ $\mu$eV}, and MADMAX and plasma haloscopes at masses $\sim$40--400 $\mu$eV~\cite{MADMAX:2019pub,Lawson:2019brd}.  However, astrophysical searches such as that presented in this work play an important role in constraining higher axion masses near and above the meV scale.
Axions with $m_a \gtrsim {\rm meV}$ are difficult to probe in the laboratory, even under the non-trivial assumption that the axion is DM (but see~\cite{Arvanitaki:2014dfa,ARIADNE:2017tdd} for a proposal).  While it was previously thought that the QCD axion cannot explain the entirety of DM at masses at and above $\sim$meV masses, this assumption has been challenged recently (see,~{\it e.g.},~\cite{Co:2020dya,Gorghetto:2020qws}), further motivating the search for meV-scale axions. 

\begin{figure*}[!t]
 \includegraphics[width = 1.0\textwidth]{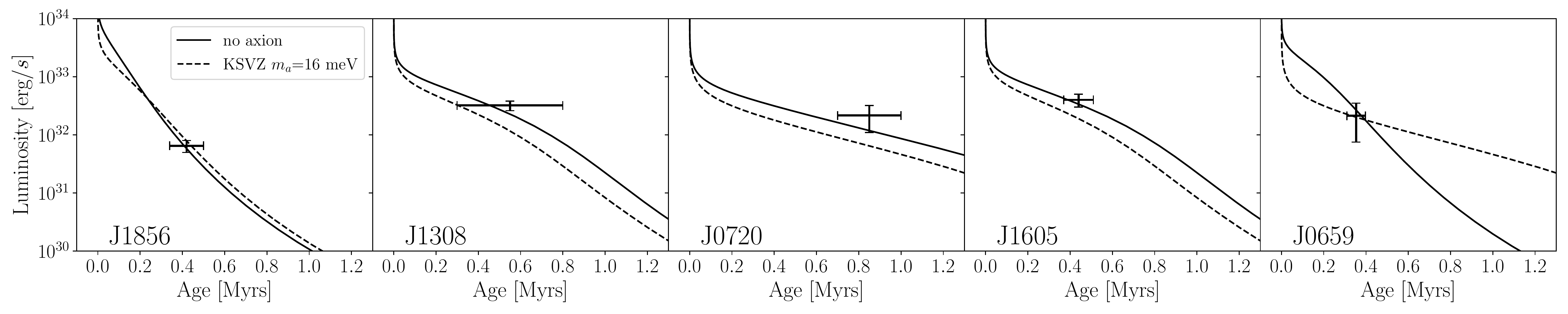}
\caption{The luminosity and age data for each of the NSs considered in this work (see Tab.~\ref{tab:M7}).  We show the best-fit cooling curves computed in this work for each of these NSs under the null hypothesis and with the axion mass fixed to $m_a = 16$ meV, which is our 95\% upper limit on the QCD axion mass in the context of the KSVZ model. 
}
\label{fig:coolingcurves}
\end{figure*}

The fundamental idea behind how axions may modify NS cooling is that these particles, just like neutrinos~\cite{Yakovlev:2000jp}, may be produced in thermal scattering processes within the NS cores and escape the stars due to their weak interactions~\cite{Iwamoto:1984ir,Iwamoto:1992jp}.  Most previous studies of axion-induced NS cooling have focused on either proto-NSs, like that from SN 1987A~\cite{Raffelt:2006cw,Fischer:2016cyd,Chang:2018rso,Carenza:2019pxu,Carenza:2020cis}, that are seconds old or young NSs like Cas A~\cite{Page:2010aw,2011MNRAS.412L.108S,Leinson:2014cja,Leinson:2014ioa,Sedrakian:2015krq,Hamaguchi:2018oqw,Leinson:2021ety}, which has an age $\sim$300 yrs.  In this work we show that robust and competitive constraints on $m_a$ may be found from analyses of older NS cooling, focusing on NSs with ages $\sim$$10^5$--$10^6$ yrs.  This is important when considering the possible issues that affect the SN 1987A and Cas A constraints, {such as the lack of fully self-consistent 3D simulations~\cite{Fischer:2016cyd} for SN 1987A and uncertainties related to the formation of the proto-NS~\cite{Bar:2019ifz}.}
Axion constraints from Cas A arise by using the observed temperature drop of the young NS over the past $\sim$two decades by the {\it Chandra} telescope, but it was realized recently that this drop may be due in large part to a systematic evolution of the energy calibration of the detector over time~\cite{Posselt:2018xaf}.  Moreover, the Cas A constraints are typically derived under the assumptions of specific superfluidity and equation of state (EOS) models, which are themselves uncertain.  While the importance of the SN 1987A and Cas A results should not be discounted, it is clear that additional, independent probes are needed to robustly disfavor or detect the QCD axion at masses above a few meV.

\begin{figure}[!t]
\includegraphics[width = 0.49\textwidth]{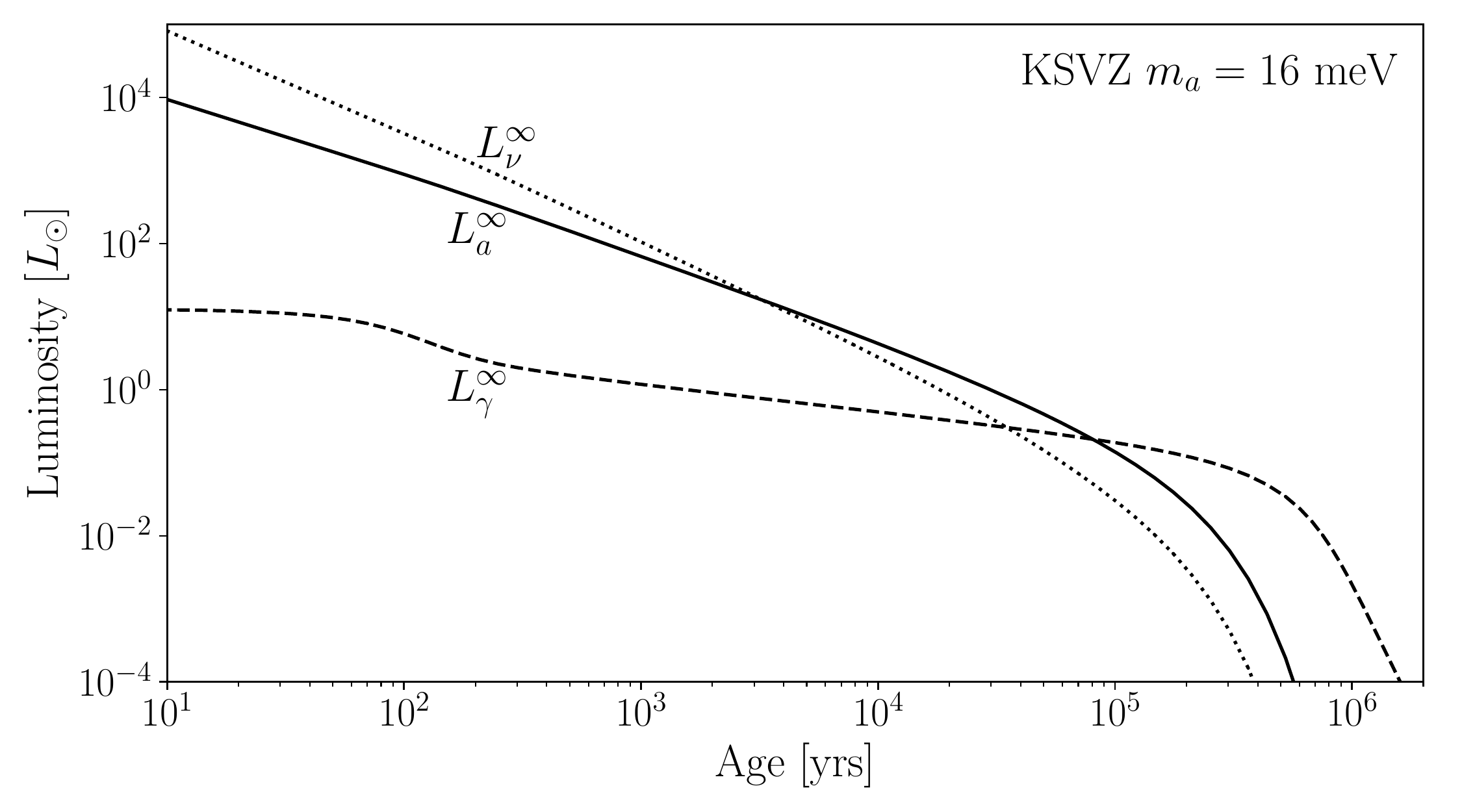}
\caption{The luminosity production from neutrinos, axions, and surface radiation for an example NS with the KSVZ axion at $m_a = 16$ meV.  The NS parameters have been chosen to be those found in the profile likelihood procedure for J1605 with this axion mass: the BSk22 EOS, \texttt{SBF-0-0} superfluidity model, $M_{\rm NS} = 1.0 $ $M_\odot$, and $\Delta M / M_\odot = 10^{-12}$. }
\label{fig:AgeLum}
\end{figure}

\noindent
{\bf Isolated NS data and modeling.---} In this work we use luminosity and kinematic age data from four of the seven Magnificent Seven (M7) NSs, which are those where kinematic age data is available (see Tab.~\ref{tab:M7} and Fig.~\ref{fig:coolingcurves} for their relevant data).  {We add to this list PSR J0659,
identified with the Monogem Ring, as it  
also has an age above $10^5$ yrs known from kinematic considerations~\cite{Potekhin:2020ttj,Suzuki:2021ium} and a thermal luminosity measurement.}
{The NSs with ages $\sim$$10^5$ yrs live at a unique era, as illustrated in Fig.~\ref{fig:AgeLum}, where cooling from axion bremsstrahlung emission is maximally important; at lower ages neutrino emission plays a more important role since the the neutrino (axion) emissivity scales as $\propto T^8$ ($T^6$) with temperature $T$, while at older ages the thermal surface emission dominates the energy loss.}
We discuss NSs with ages less than $10^5$ yrs, including Cas A, in the Supplementary Material (SM).
The age data have been determined by tracing the NSs back to their birthplaces. A measured NS orbit is run backwards in the Galactic potential and a parent stellar cluster is identified in each case. J1856 and J1308 are found to originate in the Upper Scorpius OB association~\cite{2013MNRAS.429.3517M,2009AA...497..423M}. J0720 was likely born in the Trumpler association~\cite{2011MNRAS.417..617T}. J1605 can be associated with a runaway former binary companion, which was disrupted in a supernova~\cite{2012PASA...29...98T}.

The thermal luminosity data for these NSs are measured from spectral fitting of NS surface models to the X-ray spectra. The strong magnetic fields create localized temperature inhomogeneities on the surfaces, so the total thermal luminosity is a more robust observable for our purposes since it is less affected by the temperature inhomogeneities than direct temperature measurements. 
For this reason we use the luminosity data in this work rather than surface temperature measurements~\cite{Beznogov:2021ijc}. Typically, one of a NS atmosphere model or a double-blackbody model is fit to the X-ray spectral data. For J1856, a thin partially ionized hydrogen atmosphere model suggests our lower luminosity bound $\sim$5$\times 10^{31}$ erg/s~\cite{Ho:2006uk} while a double blackbody model suggests the upper bound $\sim$8$\times 10^{31}$ erg/s~\cite{2012AA...541A..66S}. For J1308, the same models suggest (3.3 $\pm$ 0.5)$\times 10^{32}$ erg/s and 2.6$\times 10^{32}$ erg/s, respectively~\cite{2011AA...534A..74H}. For J0720, both types of models give similar luminosities $\sim$2$\times 10^{32}$ erg/s~\cite{2011MNRAS.417..617T}. 
A double blackbody fit yields the luminosity  $(4 \pm 1)\times 10^{32}$ erg/s for J1605, which we adopt in our analysis~\cite{Pires:2019qsk}. The J0659 luminosity was determined with a double blackbody model including a power law, since it emits non-thermally in hard X-rays as it is a pulsar~\cite{Zharikov:2021llh}. We assume Gaussian priors on the NS luminosities and ages that include all measurements at 1$\sigma$. Note that the M7 have previously been the subject of searches for axion-induced hard $X$-ray emission~\cite{Dessert:2019dos,Buschmann:2019pfp}.

In this work we build off of the one-dimensional NS cooling code \texttt{NSCool}~\cite{2016ascl.soft09009P} to simulate NS cooling curves with axion energy losses. \texttt{NSCool} solves the energy balance and transport equations in full General Relativity in the core and crust of the NS. 
An envelope model $T_s(T_b)$ that relates the interior and surface temperatures, $T_b$ and $T_s$ respectively, is glued to the exterior of the crust. After thermal relaxation, so that the NS has a uniform core temperature, integrating the energy balance equation over the interior of the NS leads to the cooling equation
\begin{equation}
\label{eq:cooling}
    L_\gamma^\infty = -C \dfrac{dT_b^\infty}{dt} - L_\nu^\infty - L_a^\infty + H \,,
\end{equation}
where
$L_\gamma^\infty = 4\pi R_{*,\infty}^2 (T_s^\infty)^4$
is the photon luminosity, and $t$ is time. (Throughout this Letter, the infinity superscript will indicate that the value is taken to be that as measured by a distant observer.) The heat capacity of the NS is $C$, $L_\nu^\infty$ is the neutrino luminosity, $L_a^\infty$ is the axion luminosity, and $H$ accounts for possible heating sources, such as from magnetic field decay (see Fig.~\ref{fig:AgeLum} for an illustration).
Note that we include important corrections to the neutrino emissivities relative to those in \texttt{NSCool}~\cite{2016ascl.soft09009P}, which we discuss shortly, and we also assume that $H = 0$, since magnetic field induced heating likely plays a subdominant role in constraining $L_a^\infty$ (see the SM).
The solution of this equation yields the NS cooling curve $L_\gamma^\infty(t,{\bm \theta})$, where ${\bm \theta}$ parameterizes the 
particular choices of   axion and NS properties.  The axion is parameterized by its mass $m_a$ and coupling constants to nucleons, while for the NS we need to know (i) the NS mass $M_{\rm NS}$, (ii) the equation of state (EOS), (iii) the superfluidity model $\Delta(T_b)$, and (iv) the envelope model parameterized by the mass of light elements $\Delta M$.

 The axion energy losses from nucleon scattering processes are determined by the axion-neutron and axion-proton dimensionless coupling constants $C_p$ and $C_n$, respectively, in addition to $f_a$; the axion-nucleon interactions are of the form ${\mathcal L} \supset (C_N / 2 f_a) \bar \psi_N \gamma^\mu \gamma_5 \psi_N \partial_\mu a$ with $N = p,n$, $\psi_N$ the nucleon fields, and $a$ the axion field.  In the KSVZ axion model $C_p = -0.47 \pm 0.03$ and $C_n = -0.02 \pm 0.03$~\cite{diCortona:2015ldu}, while in the DFSZ model $C_p$ and $C_n$ are functions of $\tan \beta$, which is the ratio of the vacuum expectation values of the up- to down-type Higgs doublets in that theory: $C_n = (-0.160 \pm 0.025) + 0.414 \sin^2 \beta$, $C_p = (-0.182 \pm 0.025) - 0.435 \sin^2 \beta$~\cite{diCortona:2015ldu}.  
Additional axion models are also possible~\cite{DiLuzio:2020wdo}, for which it is useful to define the dimensionless coupling constants $g_{aNN} = C_N m_N / f_a$, with $m_N$ the nucleon mass.  Note that the uncertainties on the KSVZ and DFSZ axion couplings arise from lattice QCD~\cite{diCortona:2015ldu}; to make contact with previous literature we assume the central values.

When computing the axion luminosities we account for axion bremsstrahlung~\cite{Iwamoto:1984ir,Iwamoto:1992jp} from nucleons and axion production from Cooper pair breaking and formation (PBF).  If the NS core temperature is below the superfluid critical temperature, nucleons form Cooper pairs and condense into a superfluid phase.   
These Cooper pairs can  liberate energy in the form of neutrinos~\cite{1976ApJ...205..541F,1987PhLB..184..119S} or axions~\cite{Keller:2012yr,Sedrakian:2015krq} when breaking and forming.
The PBF processes {may} dominate the axion luminosity at temperatures near the superfluid transition temperature, while the bremsstrahlung processes are exponentially suppressed at lower temperatures.
To evaluate the axion and neutrino emission rates, for both PBF and bremsstrahlung production, we account for the medium-dependent axion-nucleon and pion-nucleon couplings~\cite{Fischer:2016boc}, which have not been included in earlier work on axion emission from compact stars or supernovae.
These corrections are density dependent, varying throughout the interior of the star, and at the highest densities the overall effect is a $\sim$$30\%$ suppression of the axion emission rate and a $\sim$$50\%$ enhancement of the neutrino rate. 
See the SM for details.

\begin{table}[t]
\begin{tabular}{|c|c|c|c|c|}
\hline
Name   & $L_\gamma^\infty$ [$10^{33}$ erg/s] & Age [Myr]       & Refs                                                                                         \\ \hline
J1856             & $0.065 \pm 0.015$          & $0.42 \pm 0.08$ & \cite{Ho:2006uk,2012AA...541A..66S,2013MNRAS.429.3517M} \\ \hline
J1308           & $0.32 \pm 0.06$          & $0.55 \pm 0.25$ & \cite{2011AA...534A..74H,2009AA...497..423M}                                  \\ \hline
J0720           & $0.22 \pm 0.11$           & $0.85 \pm 0.15$ & \cite{Hambaryan:2017wvm,2011MNRAS.417..617T}                                                 \\ \hline
J1605           & $0.4 \pm 0.1$          & $0.44 \pm 0.07$ & \cite{Pires:2019qsk,2012PASA...29...98T}                                    \\ \hline
J0659           & $0.28 \pm 0.14$          & $0.35 \pm 0.044$ & \cite{Suzuki:2021ium,Zharikov:2021llh}                                    \\ \hline
\end{tabular}
\caption{\label{tab:M7} The properties of the NSs considered in this work -- RX J1856.6$-$3754, RX J1308.6$+$2127, RX  J0720.4$-$3125, RX J1605.3$+$3249, PSR J0659$+$1414  -- which we abbreviate throughout this Letter.  We include all known NSs with ages above $10^5$ yrs and robust age and luminosity measurements (see {\it e.g.}~\cite{Potekhin:2020ttj}).  Younger NSs are discussed in the SM.}
\end{table}

We make one additional modification to \texttt{NSCool} to help quantify the effects of astrophysical uncertainties.
The addition of light elements (hydrogen, helium, and carbon) in the NS envelope changes the expected relation between the surface and core temperatures, which in turn affects the observed surface luminosity even for the same internal state. We incorporate the analytic formulae in~\cite{Potekhin:1997mn} into \texttt{NSCool} in order to cool a NS with a mass $\Delta M$ of light elements layered on top of the default iron surface. Values for $\Delta M$ can span from $0\, M_\odot$, such that the NS has a pure iron surface, to $\sim$ $10^{-7} M_\odot$, which is the mass of the entire envelope. In practice, we modify the equation $T_s(T_b)$ to account for the addition of light elements, which can change the photon luminosity of the NS by up to a factor $\sim$5  before the photon cooling stage and $\gtrsim$100 after.  Since each $\Delta M$ value requires a dedicated \texttt{NSCool} simulation, we use a discrete number eight of equally log-spaced values for $\Delta M / M_\odot$ ranging from $10^{-20} M_\odot$ to $10^{-6} M_\odot$.  Similarly, we discretize the NS mass range with six equally spaced masses between $1 M_\odot$ and $2 M_\odot$; we show in the SM that our results are not strongly dependent on this mass range.

We simulate NSs for five distinct EOSs: APR~\cite{PhysRevC.58.1804}, BSk22, BSk24, BSk25, and BSk26~\cite{Pearson:2018tkr}. The APR EOS is constructed using variational methods to model the two-nucleon interaction incorporating the effects of many-body interactions and with the input of nucleon-nucleon scattering data. The BSk family of EOSs are generated by fitting the Skyrme effective interaction to atomic mass data. The distinct BSk EOSs are constructed with different assumed values of the Skyrme symmetry energy. These EOSs phenomenologically characterize the range of possible stiffnesses of the EOSs.  Recently, data from the {\it NICER} telescope has allowed for the simultaneous measurements of the mass and radius of two NSs, PSR J0030~\cite{Riley:2019yda} and PSR J0740~\cite{Miller:2021qha}, which can be used in conjunction with gravitational wave observations of NS mergers to constrain the EOS~\cite{Miller:2021qha}.  As we show in SM Fig.~\ref{fig:EOS_MR}, only the BSk22, BSk24, and BSk25 EOS are consistent with the mass-radius data to within 1--2$\sigma$ significance.  We thus restrict ourselves to this set of EOS in the main Letter, though we discuss how our results change with the APR and BSk26 EOS in the SM.

We consider three distinct superfluidity models, denoted in \texttt{NSCool} and here as \texttt{0-0-0}, \texttt{SFB-0-0}, and \texttt{SFB-0-T73}.
The first model assumes no superfluidity by setting the gaps to zero. The second model turns on the $\tensor[^1]{S}{_0}$ neutron pairing gap from~\cite{Schwenk:2002fq}, and the third model additionally turns on the $\tensor[^1]{S}{_0}$ proton pairing gap from~\cite{BALDO1992349}.  {Neutron $\tensor[^3]{P}{_2}$-$\tensor[^3]{F}{_2}$ pairing may also be possible (we will refer to this as $\tensor[^3]{P}{_2}$ for brevity)}, but the estimate of this gap is more complicated in part because it appears at higher density where many-body interactions are more important (see, {\it e.g.},~\cite{Sedrakian:2018ydt}).  However, in the SM we show that the $\tensor[^3]{P}{_2}$ superfluid would only increase the strength of our limit, though many $\tensor[^3]{P}{_2}$ gap models are inconsistent with the isolated NS data.

\noindent
{\bf Data analysis and results.---} Given the set of cooling curves, we can compare them to the observed data in Tab.~\ref{tab:M7}. For a given QCD axion model, under the assumption of a particular NS  $\mathrm{EOS}$ and superfluidity model $ \Delta(T_b)$, let us label the present-day luminosity of a NS by $L(m_a, {\bm \theta})$. The luminosity of NS $i$ is then jointly determined by the axion mass $m_a$ and the nuisance parameters ${\bm \theta^i} = \{M^i_{\rm NS},\DeltaM^i,t^i\}$ that characterize the NS.
We can now write the likelihood for a single NS $i$ as 
\es{eq:indL}{
\mathcal{L}_i({\bm d_i}|m_a,{\bm \theta^i}) = & \mathcal{N}(L(m_a,{\bm \theta}^i)-L^i_{0},\sigma_{L}^{i}) \\ &\times \mathcal{N}(t^i - t^i_0 , \sigma_{t}^{i}) \,,
}
where we have introduced the NS data set ${\bm d_i} = \{L^i_{0},\sigma_{L}^{i},t^i_{0},\sigma_{t}^{i}\}$, where $L^i_0$ is the measured luminosity of the NS with uncertainty $\sigma_{L}^{i}$.  Similarly, $t^i_{0}$ is the measured age of the NS with uncertainty $\sigma_{t}^{i}$.  The probability of observing a value $x$ under the zero-mean Gaussian distribution with standard deviation $\sigma$ is denoted by $\mathcal{N}(x,\sigma)$.
The joint likelihood $\mathcal{L}({\bm d}|m_a,{\bm \theta})$ over all five NSs is constructed by taking the product of~\eqref{eq:indL} over the  NSs.  Note that the total list of model parameters is denoted by ${\bm \theta} = \{ {\bm \theta^i}\}_{i=1}^5$. The best-fit axion mass $\hat m_a$ and nuisance parameters $\hat{\bm{\theta}}$ can be determined for a given choice of EOS and superfluidity model by maximizing the joint likelihood.  To test for systematic mismodeling we allow $m_a < 0$, with the axion luminosity multiplied by ${\rm{sign}}(m_a)$.

Additionally, given the large number of nuisance parameters, many of which have non-trivial degeneracy with the signal parameter $m_a$, we determine the 95\% upper limit on $m_a$, defined by $m_a^{\rm 95}$, by the Neyman construction of the 95\% confidence interval for $m_a$ through a Monte Carlo (MC) procedure rather than by invoking Wilks' theorem.  Similarly, we determine the significance of the axion model over the null hypothesis through MC simulations of the null hypothesis, instead of relying on Wilks' theorem.  (See the SM for details.)

\begin{figure}[!t]
\includegraphics[width = 0.49\textwidth]{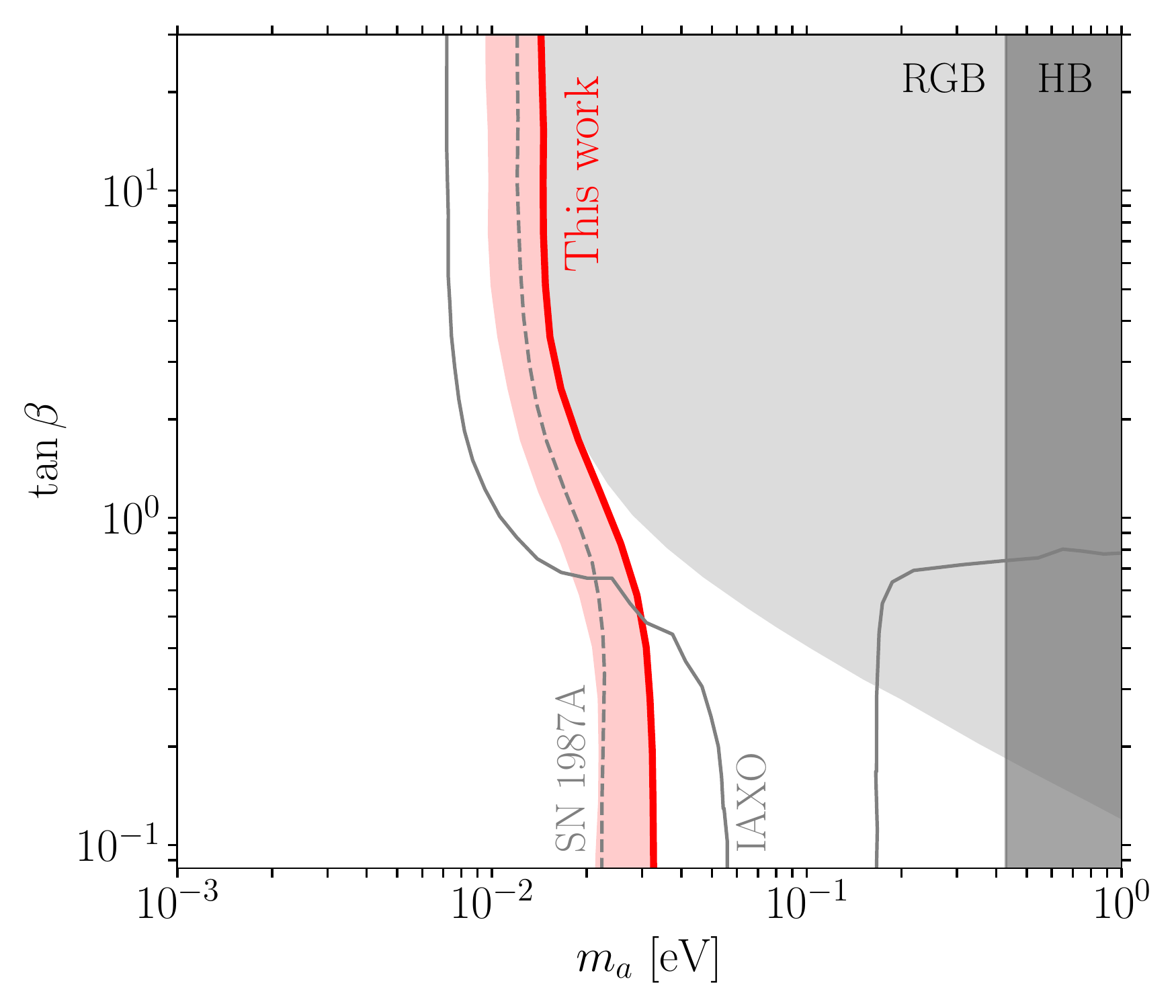}
\caption{Upper limit from this work on the DFSZ axion mass $m_a$ as a function of $\tan \beta$, which 
controls the relative coupling of the axion to neutrons and protons.  The width of the shaded red band reflects the uncertainty on the upper limit by varying over superfluidity and EOS models.  {We compare our upper limits to existing constraints and the projected IAXO discovery sensitivity.}  }
\label{fig:tanbeta}
\end{figure}

For each combination of EOS and superfluidity model we determine $m_a^{95}$, $\hat m_a$, and the significance of the axion model over the null hypothesis of $m_a = 0$ meV.
We choose the 95\% upper limit over the ensemble of nine EOS and superfluidity combinations that gives the most conservative limit.  For the KSVZ axion model we find that $m_a^{95} \approx 16$ meV with the BSk22 EOS model and the \texttt{SFB-0-0} superfluidity model; the strongest constraint over all combinations is $m_a^{95} \approx 6$ meV with the BSk25 EOS and the \texttt{SFB-0-T73} model.  With that said, the \texttt{SFB-0-T73} model is the worst fit to the data, with the best-fit axion mass being negative at $\sim$1.6$\sigma$ significance.  The best-fitting model is that with the BSk22 EOS and no superfluidity, for which the limit is $m_a^{95} \approx 14$ meV and the best-fit axion mass being negative at $\sim$0.36$\sigma$.   
From these results we conclude that the NS cooling data show no evidence for the KSVZ axion and also no significant evidence for a preference for a particular EOS or superfluidity combination; the isolated NS data appear well described by the null hypothesis.

For the DFSZ axion the results depend on the value of $\tan \beta$.  
In Fig.~\ref{fig:tanbeta} we show $m_a^{95}$ as a function of $\tan \beta$,
with the shaded band showing the range of limits found over all EOS and superfluidity combinations. DFSZ axion masses to the right of the exclusion curve are disfavored at 95\% confidence.  The weakest limit (bold) is achieved for all $\tan \beta$ for the no superfluidity model with the BSk22 EOS.
We compare these upper limits to those from horizontal branch (HB)~\cite{Ayala:2014pea,Straniero:2015nvc}, red giant branch (RGB)~\cite{Viaux:2013lha,2018arXiv180210357S}, and SN 1987A~\cite{Carenza:2019pxu} cooling.  Note, however, that the SN 1987A limit is approximate, since {\it e.g.} it arises from the rough requirement $L_a^\infty < L_\nu^\infty$ for the proto-NS, and also it does not account for the density-dependent couplings for axions and neutrinos, which we estimate should weaken the SN 1987A limit by a factor $\sim$1.3--1.6, depending on the EOS.
{We also show the projected discovery reach for the future IAXO experiment~\cite{IAXO:2019mpb}; our results leave open a narrow mass range $\sim$10 meV where IAXO may discover the QCD axion.}  
In the axion model with only an axion-neutron (axion-proton) coupling  we constrain  $|g_{ann}| < 1.3\times 10^{-9}$ ($|g_{app}| < 1.5\times 10^{-9}$) at 95\% confidence.

\noindent
{\bf Discussion.---}In this Letter we present a search for the QCD axion from NS cooling, comparing NS cooling simulations with axions to luminosity and kinematically-determined age data from five NSs.  Four of the five NSs are part of the M7, which are unique in that they only emit radiation thermally and thus have well-measured thermal luminosities.  The NSs that are most important for our analysis are J0720, J1605, and J1308, as further highlighted in the SM.      

Our upper limits disfavor at 95\% QCD axions with masses {$m_a \gtrsim 10 - 30$ meV}, depending on the axion model, which constrains the axion interpretation of the previously-observed stellar cooling anomalies~\cite{Giannotti:2017hny}.  The limits may be stronger if $\tensor[^3]{P}{_2}$ superfluidity is active in the NS cores, as we discuss in the SM, though large $\tensor[^3]{P}{_2}$ gaps appear disfavored by the isolated NS data.
Many-body nuclear techniques should provide improved estimates of the energy gaps of the $\tensor[^1]{S}{_0}$ (neutron), $\tensor[^1]{S}{_0}$ (proton), and $\tensor[^3]{P}{_2}$ (neutron) pairings in the future~\cite{Sedrakian:2018ydt}.  On the other hand, more work should be done to rigorously assess the possible impact of heating mechanisms such as magnetic field decay on the axion limits, for example using fully self-consistent simulations along the lines of those in~\cite{Aguilera:2007xk,Potekhin:2017ufy}.   Axions may also be produced from more exotic forms of matter in the NS interiors, such as hyperon superfluids and pionic and kaonic Bose Einstein condensates, and these channels should be investigated as the NS EOS and composition becomes better understood.

\begin{acknowledgments}
{\it 
We thank K. Cranmer, S. McDermott, T. Opferkuch, G. Raffelt, and S. Witte for useful discussions and T. Opferkuch for sharing data related to the EOS. 
M.B. was supported by
the DOE under Award Number DESC0007968.
A.J.L. was supported in part by the National Science Foundation under Award No. 2114024. 
C.D. and B.R.S. were supported  in  part  by  the  DOE  Early Career  Grant  DESC0019225. JF was supported by a Pappalardo Fellowship. This research used resources from the Lawrencium computational cluster provided by the IT Division at the Lawrence Berkeley National Laboratory, supported by the Director, Office of Science, and Office of Basic Energy Sciences, of the U.S. Department of Energy under Contract No.  DE-AC02-05CH11231.
}

\end{acknowledgments}

\bibliography{axion_cyclotron}

\clearpage

\onecolumngrid
\begin{center}
  \textbf{\large Supplementary Material for Upper Limit on the QCD Axion Mass from Isolated Neutron Star Cooling}\\[.2cm]
  \vspace{0.05in}
  {Malte Buschmann, Christopher Dessert, Joshua W. Foster, Andrew J. Long, and Benjamin R. Safdi}
\end{center}

\twocolumngrid
\setcounter{equation}{0}
\setcounter{figure}{0}
\setcounter{table}{0}
\setcounter{section}{0}
\setcounter{page}{1}
\makeatletter
\renewcommand{\theequation}{S\arabic{equation}}
\renewcommand{\thefigure}{S\arabic{figure}}
\renewcommand{\thetable}{S\arabic{table}}

\onecolumngrid

This Supplementary Material (SM) is organized as follows.  In Sec.~\ref{sec:axion_rates} we present our calculations of the axion and neutrino emissivities.  
Sec.~\ref{app:stat} presents our statistical methodology.  Sec.~\ref{app:extended} gives extended results for the analyses mentioned in the main body, while Sec.~\ref{app:B-field} presents our estimates for the effects of magnetic field decay on the axion upper limits.

\section{Axion and neutrino emissivities}
\label{sec:axion_rates}

In this section we present the axion and neutrino emissivities that we use in the simulations discussed in the main Letter.  We include a number of factors relevant for axion and neutrino production in dense media that have not previously been included in NS cooling simulations.

\subsection{Axion production rates}

Our analysis accounts for the two dominant channels by which axions are produced in the core of a cooling NS.  
When the core temperature $T$ exceeds the critical temperature $T_c$ for the superfluid phase transition, axion emission is dominated by nucleon-nucleon bremsstrahlung.  
When $T$ falls below $T_c$, axion emission is dominated by the formation and breaking of Cooper pairs (PBF processes).  
To calculate the axion production rate, we sum these two contributions.  

Axion emission via nucleon-nucleon bremsstrahlung corresponds to three scattering channels: $nn \to nna$, $pp \to ppa$, and $np \to npa$.  
For the temperatures of interest, the nucleons are strongly degenerate and non-relativistic.  
Expressions for the axion emissivity $\varepsilon_a$ (energy emitted per volume per time) are provided by Refs.~\cite{Iwamoto:1984ir,Iwamoto:1992jp}.  
These early derivations of the axion emissivity did not take into account various medium effects, which were pointed out in later literature, and which we have incorporated into the calculation.  
The axion emissivities that we use in our work are as follows: 
\begin{align}
    nn \to nna: & \qquad 
	\varepsilon_a \simeq 
	\bigl( 7.373 \times 10^{11} \ \mathrm{erg} / \mathrm{cm}^3 / \mathrm{sec} \bigr) 
	\biggl( \frac{g_{ann}}{10^{-10}} \biggr)^2 
	\biggl( \frac{F(x_n)}{0.601566} \biggr) 
	\biggl( \frac{p_{F,n}}{1.68 \ \mathrm{fm}^{-1}} \biggr) 
	\biggl( \frac{T}{10^8 \ \mathrm{K}} \biggr)^6 
	\\ & \hspace{5.5cm} \times 
	\biggl( \frac{\beta_{nn}}{0.56} \biggr) 
	\biggl( \frac{\gamma_{nn}}{0.838} \biggr) 
	\biggl( \frac{\gamma}{1} \biggr)^6 
	\biggl( \frac{\mathcal{R}_{nn}}{1} \biggr) \nonumber \\ 
    pp \to ppa: & \qquad 
	\varepsilon_a \simeq 
	\bigl( 9.191 \times 10^{11} \ \mathrm{erg} / \mathrm{cm}^3 / \mathrm{sec} \bigr) 
	\biggl( \frac{g_{app}}{10^{-10}} \biggr)^2 
	\biggl( \frac{F(x_p)}{0.601556} \biggr) 
	\biggl( \frac{p_{F,p}}{1.68 \ \mathrm{fm}^{-1}} \biggr) 
	\biggl( \frac{T}{10^8 \ \mathrm{K}} \biggr)^6 
	\\ & \hspace{5.5cm} \times 
	\biggl( \frac{\beta_{pp}}{0.7} \biggr) 
	\biggl( \frac{\gamma_{pp}}{0.838} \biggr) 
	\biggl( \frac{\gamma}{1} \biggr)^6 
	\biggl( \frac{\mathcal{R}_{pp}}{1} \biggr) \nonumber \\ 
    np \to npa: & \qquad 
	\varepsilon_a \simeq 
	\bigl( 9.617 \times 10^{11} \ \mathrm{erg} / \mathrm{cm}^3 / \mathrm{sec} \bigr) 
	\biggl( \frac{g_\mathrm{eff}}{10^{-10}} \biggr)^2 
	\biggl( \frac{p_{F,p}}{1.68 \ \mathrm{fm}^{-1}} \biggr) 
	\biggl( \frac{T}{10^8 \ \mathrm{K}} \biggr)^6 
	\\ & \hspace{5.5cm} \times 
	\biggl( \frac{\beta_{np}}{0.66} \biggr) 
	\biggl( \frac{\gamma_{np}}{0.838} \biggr) 
	\biggl( \frac{\gamma}{1} \biggr)^6 
	\biggl( \frac{\mathcal{R}_{np}}{1} \biggr) \nonumber 
	\;.
\end{align}
Note that $1.68 \ \mathrm{fm}^{-1} \simeq 331.5 \ \mathrm{MeV}$, $10^8 \ \mathrm{K} \simeq 8.617 \ \mathrm{keV}$, $m_n \simeq 939.565 \ \mathrm{MeV}$, and $m_p \simeq 938.272 \ \mathrm{MeV}$. 
 
Let us now discuss each of the factors appearing in the emissivities above.  
\begin{itemize}
    \item  The axion emissivity is proportional to $\gann^2$ ($\gapp^2$) if the axion couples to a neutron (proton) only.  If the axion couples to both nucleons, then the axion emissivity for the process $n p \to n p a$ depends on an effective coupling~\cite{Iwamoto:1992jp}
\begin{equation}
\begin{split}
    g_\mathrm{eff} & = \sqrt{(g_{app} + g_{ann})^2 \, C_g + (g_{app} - g_{ann})^2 \, C_h} \\ 
	C_g & = 
	\frac{1}{2} F(x_p) 
	+ F\Bigl(\frac{2x_nx_p}{x_n+x_p}\Bigr)
	+ F\Bigl(\frac{2x_nx_p}{x_p-x_n}\Bigr)
	+ \frac{x_p}{x_n} F\Bigl(\frac{2x_nx_p}{x_n+x_p}\Bigr)
	- \frac{x_p}{x_n} F\Bigl(\frac{2x_nx_p}{x_p-x_n}\Bigr) 
	+ G(x_p) 
	\\ 
	C_h & = 
	\frac{1}{2} F(x_p) 
	+ \frac{1}{2} F\Bigl(\frac{2x_nx_p}{x_n+x_p}\Bigr)
	+ \frac{1}{2} F\Bigl(\frac{2x_nx_p}{x_p-x_n}\Bigr)
	+ \frac{1}{2} \frac{x_p}{x_n} F\Bigl(\frac{2x_nx_p}{x_n+x_p}\Bigr)
	- \frac{1}{2} \frac{x_p}{x_n} F\Bigl(\frac{2x_nx_p}{x_p-x_n}\Bigr) 
	+ G(x_p) 
	\\
	F(x) & = 
	1 - \frac{3}{2} \, x \, \mathrm{arctan} \frac{1}{x} + \frac{1}{2} \frac{x^2}{1+x^2} 
	\\
    G(x) & = 
    1 - x \, \mathrm{arctan} \frac{1}{x} 
    \\
    x_n & = m_{\pi^\pm} / 2 p_{F,n} \simeq 0.207 \, (p_{F,n} / 1.68 \ \mathrm{fm}^{-1})^{-1} 
    \\ 
    x_p & = m_{\pi^\pm} / 2 p_{F,p} \simeq 0.207 \, (p_{F,p} / 1.68 \ \mathrm{fm}^{-1})^{-1} 
    \\ 
    x_e & = m_{\pi^\pm} / 2 p_{F,e} \simeq 0.207 \, (p_{F,e} / 1.68 \ \mathrm{fm}^{-1})^{-1} 
    \;.
\end{split}
\end{equation}
    \item  The dependence on the nucleon Fermi momenta, $p_{F,n}$ and $p_{F,p}$, are identical to Refs.~\cite{Iwamoto:1984ir,Iwamoto:1992jp}.  Similarly the temperature dependence is identical.  Both protons and neutrons are assumed to have the same temperature $T$.  
    \item  We add the factors of $\beta_{nn} = 0.56$, $\beta_{pp} = 0.7$, and $\beta_{np} = 0.66$.  These factors account for short-range correlations induced by the hard core of the nucleon-nucleon interactions.  The nuclei interact by pion exchange, which corresponds to a Yukawa potential $V(r)$, but the potential is suppressed at separations smaller than the nucleon radius.  In the context of neutrino emission, this effect was discussed in Refs.~\cite{Friman:1979ecl,1995AA...297..717Y}, which also provide the numerical values that we use.  
    \item  We add the factors of $\gamma_{nn} = \gamma_{pp} = \gamma_{np} = 0.838$.  The emissivities provided by Refs.~\cite{Iwamoto:1984ir,Iwamoto:1992jp} are derived under the one-pion exchange approximation (OPE).  Graphs with multiple pion exchanges can suppress the matrix element through a destructive interference.  Following Ref.~\cite{Carenza:2019pxu}, we account for two-pion exchange with an effective one-meson exchange.  Provided that the temperature is $T < O(10 \ \mathrm{erg})$ and the momentum transfer is small compared to the pion mass, then the squared amplitude is suppressed by a momentum-independent factor of $\gamma_{nn} \approx \gamma_{pp} \approx \gamma_{np} \approx (1 - C_\rho m_{\pi^0}^2 / m_\rho^2)^2 \simeq 0.838$ for $C_\rho = 1.67$, $m_{\pi^0} \simeq 134.976 \ \mathrm{erg}$ and $m_\rho = 600 \ \mathrm{erg}$.  
    \item  We account for a suppression of the nucleon couplings in high-density NS matter as compared to their values in vacuum. To see how this suppression arises, we first substitute $f \to f^\ast$ and $g_{aNN} \to g_{aNN}^\ast$ in the expressions from Refs.~\cite{Iwamoto:1984ir,Iwamoto:1992jp} to indicate that these are in-medium couplings.  Then 
\begin{equation}
\begin{split}
    \biggl( \frac{f^\ast}{m_\pi} \biggr)^4 \bigl( g_{aNN}^\ast \bigr)^2 \bigl( m_N^\ast \bigr)^2 
    \approx \frac{ g_{\pi NN}^4 g_{aNN}^2}{16 m_N^2} \biggl( \frac{g_{\pi NN}^\ast}{g_{\pi NN}} \biggr)^4 \biggl( \frac{g_{aNN}^\ast}{g_{aNN}} \biggr)^2 \biggl( \frac{m_N^\ast}{m_N} \biggr)^{-2} 
    \approx \frac{g_{\pi NN}^4 g_{aNN}^2}{16 m_N^2} \gamma^6
    \;.
\end{split}
\end{equation}
In the first equality we have used $f^\ast / m_\pi \approx g_{\pi NN}^\ast / 2 m_N^\ast$.  
In the second equality we have used the scaling laws from \cite{Mayle:1989yx}: the Goldberger-Tremain relation $(g_{\pi NN}^\ast/g_{\pi NN}) \approx (m_N^\ast / m_N) (g_A^\ast / g_A) (f_\pi^\ast / f_\pi)^{-1}$, the Brown-Rho scaling relation $(m_N^\ast / m_N) \approx (f_\pi^\ast / f_\pi)$, and the Mayle scaling relation $(g_{aNN}^\ast / g_{aNN}) \approx (m_N^\ast / m_N) (g_A^\ast / g_A)$.  
Then $\gamma \approx (g_A^\ast / g_A)$ is given by~\cite{Fischer:2016boc} 
\begin{align}\label{eq:gamma_def}
    \gamma = \left[ 1 + \frac{1}{3} \biggl( \frac{m_N^\ast}{m_N} \biggr) \biggl( \frac{p_{F,n}}{1.68 \ \mathrm{fm}^{-1}} \biggr) \right]^{-1} 
    \;.
\end{align}
Note that $0 < \gamma < 1$, such that $\gamma \to 1$ for low-density NS matter, whereas for neutron densities a few times larger than the nuclear saturation density we have $\gamma = O(0.1)$.  We neglect medium-dependent corrections to the pion mass, which could potentially enhance the axion emission rate \cite{Mayle:1989yx}.  The net effect of accounting for the medium-dependent couplings is to introduce a factor of $(m_N^\ast/m_N)^{-2} \gamma^6$.  
    \item  We add the factors of $\mathcal{R}_{nn}$, $\mathcal{R}_{pp}$, and $\mathcal{R}_{np}$ to account for a suppression of the bremstrahlung rates due to superfluidity~\cite{1994AstL...20...43L,1995AA...297..717Y}.  When the NS core temperature falls below the superfluid critical temperature, $T < T_c$, nucleons can form Cooper pairs as the system partially condenses into a baryonic superfluid.  As nucleons are bound into Cooper pairs, there are fewer free nucleons, which suppresses the bremsstrahlung rate exponentially.  
\end{itemize}

If the temperature is not far below the critical temperature, Cooper pairs can also be broken by scattering.  
The formation and breaking of Cooper pairs can produce axions and neutrinos that carry away the liberated binding energy.  
If the NS matter is in the superfluid phase, these PBF processes provide the dominant axion production channels~\cite{Yakovlev:1998wr}. 

Phases of baryonic superfluids can be distinguished by the spin and flavor of the paired nucleons.  
In this work, we consider the spin-$0$ $S$-wave neutron-neutron pairing, the spin-$0$ $S$-wave proton-proton pairing, and the spin-$1$ $P$-wave neutron-neutron pairing (SM only).  
{
We do not consider the neutron-proton $D$-wave pairing, which is easily disrupted by a small difference between the proton and neutron densities~\cite{Haskell:2017lkl}.  
}
Each pairing has an associated energy gap in the quasiparticle spectrum, called the superfluid pairing gap $\Delta$.  
For the neutron $P$-wave superfluidity, there are two possible types of pairings, called $P^A$-wave and $P^B$-wave, which differ in the anisotropy of the gap energy~\cite{1994AstL...20...43L,1994ARep...38..247L,Yakovlev:1998wr}. 
We only include the $P^A$-wave pairing in our analysis, since the results are similar for the $P^B$-wave pairing. 
{
The temperature dependence of the pairing gaps are provided by \cite{Yakovlev:1998wr}, and the corresponding superfluid critical temperatures are provided by \cite{Page:2004fy}.  
We do not account for uncertainties in the pairing gaps when deriving limits on the axion parameters.  
We find that our limits change by only a few percent between a model with no superfluidity and our fiducial model for $S$-wave superfluidity.  
Since this is small compared to other sources of uncertainty in our analysis, we do not expect an uncertainty in $S$-wave pairing gaps to have a significant impact on our results.  
}

To determine whether neutron or protons form a superfluid at a given point within the star, we use the local densities to evaluate the corresponding superfluid critical temperatures \cite{Page:2004fy}.  
For protons, if the local temperature is below the critical temperature for $S$-wave superfluidity, we say that a proton superfluid is present.  
For neutrons, we perform a similar comparison using the larger of the critical temperatures for the $S$-wave and $P$-wave pairings.

We evaluate the axion emissivity $\varepsilon_a$ for each pairing. 
Expressions for the emissivities are provided by Refs.~\cite{Keller:2012yr,Sedrakian:2015krq,Buschmann:2019pfp}.  
We modify these expressions to account for the medium-dependent couplings, which introduces a factor of $(m_N^\ast/m_N)^{2} \gamma^2$.  
Thus, the axion emissivities used in our work are 
\begin{align}
	\text{PBF in $\tensor[^1]{S}{_0}(n)$}: & \qquad \varepsilon_a \simeq 
	\bigl( 4.692 \times 10^{12} \ \mathrm{erg} / \mathrm{cm}^3 / \mathrm{sec} \bigr) 
	\biggl( \frac{g_{ann}}{10^{-10}} \biggr)^2 
	\biggl( \frac{p_{F,n}}{1.68 \ \mathrm{fm}^{-1}} \biggr)^3 
	\biggl( \frac{T}{10^8 \ \mathrm{K}} \biggr)^5 
	\\ & \hspace{5.5cm} \times 
	\biggl( \frac{m_n^\ast}{m_n} \biggr)^{-1} 
	\biggl( \frac{\gamma}{1} \biggr)^2 
	\biggl( \frac{I_{a,n}^S}{0.022} \biggr) \nonumber \\  
	\text{PBF in $\tensor[^1]{S}{_0}(p)$}: & \qquad \varepsilon_a \simeq 
	\bigl( 4.711 \times 10^{12} \ \mathrm{erg} / \mathrm{cm}^3 / \mathrm{sec} \bigr) 
	\biggl( \frac{g_{app}}{10^{-10}} \biggr)^2 
	\biggl( \frac{p_{F,p}}{1.68 \ \mathrm{fm}^{-1}} \biggr)^3 
	\biggl( \frac{T}{10^8 \ \mathrm{K}} \biggr)^5 
	\\ & \hspace{5.5cm} \times 
	\biggl( \frac{m_p^\ast}{m_p} \biggr)^{-1}
	\biggl( \frac{\gamma}{1} \biggr)^2 
	\biggl( \frac{I_{a,p}^S}{0.022} \biggr) \nonumber \\  
	\text{PBF in $\tensor[^3]{P}{}_2^A(n)$}: & \qquad \varepsilon_a \simeq 
	\bigl( 3.769 \times 10^{13} \ \mathrm{erg} / \mathrm{cm}^3 / \mathrm{sec} \bigr) 
	\biggl( \frac{g_{ann}}{10^{-10}} \biggr)^2 
	\biggl( \frac{p_{F,n}}{1.68 \ \mathrm{fm}^{-1}} \biggr) 
	\biggl( \frac{T}{10^8 \ \mathrm{K}} \biggr)^5 
	\\ & \hspace{5.5cm} \times 
	\biggl( \frac{m_n^\ast}{m_n} \biggr) 
	\biggl( \frac{\gamma}{1} \biggr)^2 
	\biggl( \frac{I_{a,n}^{P^A}}{0.022} \biggr) \nonumber \\  
	\text{PBF in $\tensor[^3]{P}{}_2^B(n)$}: & \qquad \varepsilon_a \simeq 
	\bigl( 3.769 \times 10^{13} \ \mathrm{erg} / \mathrm{cm}^3 / \mathrm{sec} \bigr) 
	\biggl( \frac{g_{ann}}{10^{-10}} \biggr)^2 
	\biggl( \frac{p_{F,n}}{1.68 \ \mathrm{fm}^{-1}} \biggr) 
	\biggl( \frac{T}{10^8 \ \mathrm{K}} \biggr)^5 
	\\ & \hspace{5.5cm} \times 
	\biggl( \frac{m_n^\ast}{m_n} \biggr) 
	\biggl( \frac{\gamma}{1} \biggr)^2 
	\biggl( \frac{I_{a,n}^{P^B}}{0.022} \biggr) \nonumber 
	\;,
\end{align}
where the temperature-dependent integrals $I_{aN}^S$ are evaluated numerically and appear in Ref.~\cite{Sedrakian:2015krq}.  
If both neutrons and protons are in a superfluid phase, we sum the two emissivities.  
At low temperature, $T \ll \Delta(T)$, the emissivity is exponentially suppressed $\propto e^{-2 \Delta(T)/T}$, since most nucleons settle into stable Cooper pairs, whereas axion emission requires the formation and breaking of pairs.  

In the next subsection we present our similar modifications to the neutrino emissivities and discuss the quantitative effects of these corrections relative to previous works.

\subsection{Neutrino production rates}
\label{sec:neutrino_rates}

Neutrino emission from NS matter results from the direct URCA, modified URCA (MURCA) processes (both $n$-branch and $p$-branch), nucleon-nucleon bremsstrahlung, and PBF processes. 
{
The direct URCA processes correspond to the reactions $n \to p + e + \bar{\nu}_e$ and $p + e \to n + \nu_e$.  
The MURCA processes in the $N = n,p$ branch are $n + N \to N + p + e + \bar{\nu}_e$ and $N + p + e \to n + N + \nu_e$.
}
(Note that we do not modify the direct URCA rates from those in \texttt{NSCool} because they are less relevant to this work, since the direct URCA process, which turns on at high NS masses, causes the NSs to cool too rapidly to explain the isolated NS luminosity data.)  
The corresponding emissivities are provided by Refs.~\cite{Friman:1979ecl,Levenfish:1996bw,Yakovlev:1998wr,Yakovlev:2000jp}.  
We reproduce these formulas here for completeness, and we update these formulas to account for the medium-dependent couplings {in the same way that we did for the axion emissivities}.  This introduces a factor of $\gamma^6 (m_N^\ast/m_N)^{-4}$ for the MURCA and bremsstrahlung rates, and it introduces a factor of $\gamma^2$ for the PBF rates; see \eqref{eq:gamma_def}.  
{
We should emphasize that the same approximations were used to derive the neutrino emissivities as well as the axion emissivities above, and the nuclear physics factors are treated in the same way.
}
The formulas are summarized by 
\begin{align}
    \text{MURCA($n$):} & \qquad 
    \varepsilon_\nu \simeq 
    \bigl( 4.64 \times 10^{13} \ \mathrm{erg} / \mathrm{cm}^3 / \mathrm{sec} \bigr) 
    \biggl( \frac{p_{F,e}}{1.68 \ \mathrm{fm}^{-1}} + \frac{p_{F,\mu}}{1.68 \ \mathrm{fm}^{-1}} \biggr) 
    \biggl( \frac{T}{10^8 \ \mathrm{K}} \biggr)^8 
	\\ & \hspace{5.5cm} \times
    \biggl( \frac{\alpha_{n}}{1} \biggr) 
    \biggl( \frac{\beta_{n}}{0.68} \biggr) 
    \biggl( \frac{\gamma_{n}}{0.838} \biggr) 
    \biggl( \frac{\gamma}{1} \biggr)^6 
    \biggl( \frac{\mathcal{R}_{n}}{1} \biggr) \nonumber \\
    \text{MURCA($p$):} & \qquad 
    \varepsilon_\nu \simeq 
    \bigl( 4.62 \times 10^{13} \ \mathrm{erg} / \mathrm{cm}^3 / \mathrm{sec} \bigr) 
    \biggl( \frac{p_{F,e}}{1.68 \ \mathrm{fm}^{-1}} \biggr) 
    \biggl( \frac{(p_{F,e} + 3 p_{F,p} - p_{F,n})^2}{8 p_{F,e} p_{F,p}} \biggr) 
    \biggl( \frac{T}{10^8 \ \mathrm{K}} \biggr)^8 
	\\ & \hspace{5.5cm} \times
    \biggl( \frac{\alpha_{p}}{1} \biggr) 
    \biggl( \frac{\beta_{p}}{0.68} \biggr) 
    \biggl( \frac{\gamma_{p}}{0.838} \biggr) 
    \biggl( \frac{\gamma}{1} \biggr)^6 
    \biggl( \frac{\mathcal{R}_{p}}{1} \biggr) 
    \Theta(p_{F,e} + 3 p_{F,p} - p_{F,n}) \nonumber \\ 
    nn \to nn\nu\bar{\nu}: & \qquad 
    \varepsilon_\nu \simeq 
    \bigl( 9.18 \times 10^{11} \ \mathrm{erg} / \mathrm{cm}^3 / \mathrm{sec} \bigr) 
    \biggl( \frac{p_{F,n}}{1.68 \ \mathrm{fm}^{-1}} \biggr) 
    \biggl( \frac{T}{10^8 \ \mathrm{K}} \biggr)^8 
	\\ & \hspace{5.5cm} \times
    \biggl( \frac{\alpha_{nn}}{1} \biggr) 
    \biggl( \frac{\beta_{nn}}{0.56} \biggr) 
    \biggl( \frac{\gamma_{nn}}{0.838} \biggr) 
    \biggl( \frac{\gamma}{1} \biggr)^6 
    \biggl( \frac{\mathcal{R}_{nn}}{1} \biggr) \nonumber \\ 
    np \to np\nu\bar{\nu}: & \qquad 
    \varepsilon_\nu \simeq 
    \bigl( 2.16 \times 10^{12} \ \mathrm{erg} / \mathrm{cm}^3 / \mathrm{sec} \bigr) 
    \biggl( \frac{p_{F,p}}{1.68 \ \mathrm{fm}^{-1}} \biggr) 
    \biggl( \frac{T}{10^8 \ \mathrm{K}} \biggr)^8 
	\\ & \hspace{5.5cm} \times
    \biggl( \frac{\alpha_{np}}{1} \biggr) 
    \biggl( \frac{\beta_{np}}{0.66} \biggr) 
    \biggl( \frac{\gamma_{np}}{0.838} \biggr) 
    \biggl( \frac{\gamma}{1} \biggr)^6 
    \biggl( \frac{\mathcal{R}_{np}}{1} \biggr)  \nonumber \\ 
    pp \to pp\nu\bar{\nu}: & \qquad 
    \varepsilon_\nu \simeq 
    \bigl( 1.14 \times 10^{12} \ \mathrm{erg} / \mathrm{cm}^3 / \mathrm{sec} \bigr) 
    \biggl( \frac{p_{F,p}}{1.68 \ \mathrm{fm}^{-1}} \biggr) 
    \biggl( \frac{T}{10^8 \ \mathrm{K}} \biggr)^8 
	\\ & \hspace{5.5cm} \times
    \biggl( \frac{\alpha_{pp}}{1} \biggr) 
    \biggl( \frac{\beta_{pp}}{0.7} \biggr) 
    \biggl( \frac{\gamma_{pp}}{0.838} \biggr) 
    \biggl( \frac{\gamma}{1} \biggr)^6 
    \biggl( \frac{\mathcal{R}_{pp}}{1} \biggr) \nonumber \\ 
	\text{PBF in $\tensor[^1]{S}{}_0(n)$}: & \qquad 
	\varepsilon_\nu \simeq 
    \bigl( 1.24 \times 10^{14} \ \mathrm{erg} / \mathrm{cm}^3 / \mathrm{sec} \bigr) 
    \biggl( \frac{m_n^\ast}{m_n} \biggr)
    \biggl( \frac{p_{F,n}}{1.68 \ \mathrm{fm}^{-1}} \biggr) 
    \biggl( \frac{T}{10^8 \ \mathrm{K}} \biggr)^7 
    \biggl( \frac{\gamma}{1} \biggr)^2 
    \biggl( \frac{a_{n,s}}{1} \biggr) 
    \biggl( \frac{F_A(v)}{1} \biggr) \\
	\text{PBF in $\tensor[^1]{S}{}_0(p)$}: & \qquad 
    \varepsilon_\nu \simeq 
    \bigl( 1.24 \times 10^{14} \ \mathrm{erg} / \mathrm{cm}^3 / \mathrm{sec} \bigr) 
    \biggl( \frac{m_p^\ast}{m_p} \biggr)
    \biggl( \frac{p_{F,p}}{1.68 \ \mathrm{fm}^{-1}} \biggr) 
    \biggl( \frac{T}{10^8 \ \mathrm{K}} \biggr)^7 
    \biggl( \frac{\gamma}{1} \biggr)^2 
    \biggl( \frac{a_{p,s}}{1} \biggr) 
    \biggl( \frac{F_A(v)}{1} \biggr) \\
    \text{PBF in $\tensor[^3]{P}{}_2^A(n)$}: & \qquad 
    \varepsilon_\nu \simeq 
    \bigl( 1.24 \times 10^{14} \ \mathrm{erg} / \mathrm{cm}^3 / \mathrm{sec} \bigr) 
    \biggl( \frac{m_n^\ast}{m_n} \biggr)
    \biggl( \frac{p_{F,n}}{1.68 \ \mathrm{fm}^{-1}} \biggr) 
    \biggl( \frac{T}{10^8 \ \mathrm{K}} \biggr)^7 
    \biggl( \frac{\gamma}{1} \biggr)^2 
    \biggl( \frac{a_{n,t}}{1} \biggr) 
    \biggl( \frac{F_B(v)}{1} \biggr) \\
    \text{PBF in $\tensor[^3]{P}{}_2^B(n)$}: & \qquad 
    \varepsilon_\nu \simeq 
    \bigl( 1.24 \times 10^{14} \ \mathrm{erg} / \mathrm{cm}^3 / \mathrm{sec} \bigr) 
    \biggl( \frac{m_n^\ast}{m_n} \biggr)
    \biggl( \frac{p_{F,n}}{1.68 \ \mathrm{fm}^{-1}} \biggr) 
    \biggl( \frac{T}{10^8 \ \mathrm{K}} \biggr)^7 
    \biggl( \frac{\gamma}{1} \biggr)^2 
    \biggl( \frac{a_{n,t}}{1} \biggr) 
    \biggl( \frac{F_C(v)}{1} \biggr) 
    \;,
\end{align}
where the $\alpha$ factors are given by~\cite{Friman:1979ecl} 
\begin{align}
    \alpha_{n} \approx \alpha_p 
    & \approx \frac{2}{(1 + 4 x_n^2)^2} + \frac{2 C}{1 + 4 x_n^2} + 3 C^2 \\ 
    \alpha_{nn} 
    & \approx F(x_n) \\ 
    \alpha_{np} 
    & \approx F(x_e) + \frac{2}{(1 + 4 x_n^2)^2} + \frac{4 C}{1 + 4 x_n^2} + 6 C^2 \\ 
    \alpha_{pp}
    & \approx F(x_p) 
    \;,
\end{align}
with $C \simeq -0.157$.  The $a$ factors are given by~\cite{Page:2009fu}
\begin{align}
    a_{n,s} & \approx g_A^2 \, v_{F,n}^2 \Bigl[ \bigl( m_n^\ast / m_n \bigr)^2 + 11/42 \Bigr] \\ 
    a_{p,s} & \approx g_A^2 \, v_{F,p}^2 \Bigl[ \bigl( m_p^\ast / m_p \bigr)^2 + 11/42 \Bigr] \\ 
    a_{n,t} & \approx 2 g_A^2 
    \;,
\end{align}
with $g_A \simeq 1.26$ and $v_{F,N} = p_{F,N} / m_N^\ast$, and the $F(v)$ factors are given by Ref.~\cite{Yakovlev:1998wr}.  
For the $p$-branch MURCA emissivity, we include momentum-dependent factor from Ref.~\cite{Yakovlev:2000jp}, which corrects the factor from Ref.~\cite{Levenfish:1996bw}.  
We set $\beta_{n} = \beta_{p} = 0.68$, $\beta_{nn} = 0.56$, $\beta_{np} = 0.66$, $\beta_{pp} = 0.7$, and $\gamma_{n} = \gamma_{p} = \gamma_{nn} = \gamma_{np} = \gamma_{pp} = 0.838$.  

\begin{figure*}[!t]
 \includegraphics[width = 0.49\textwidth]{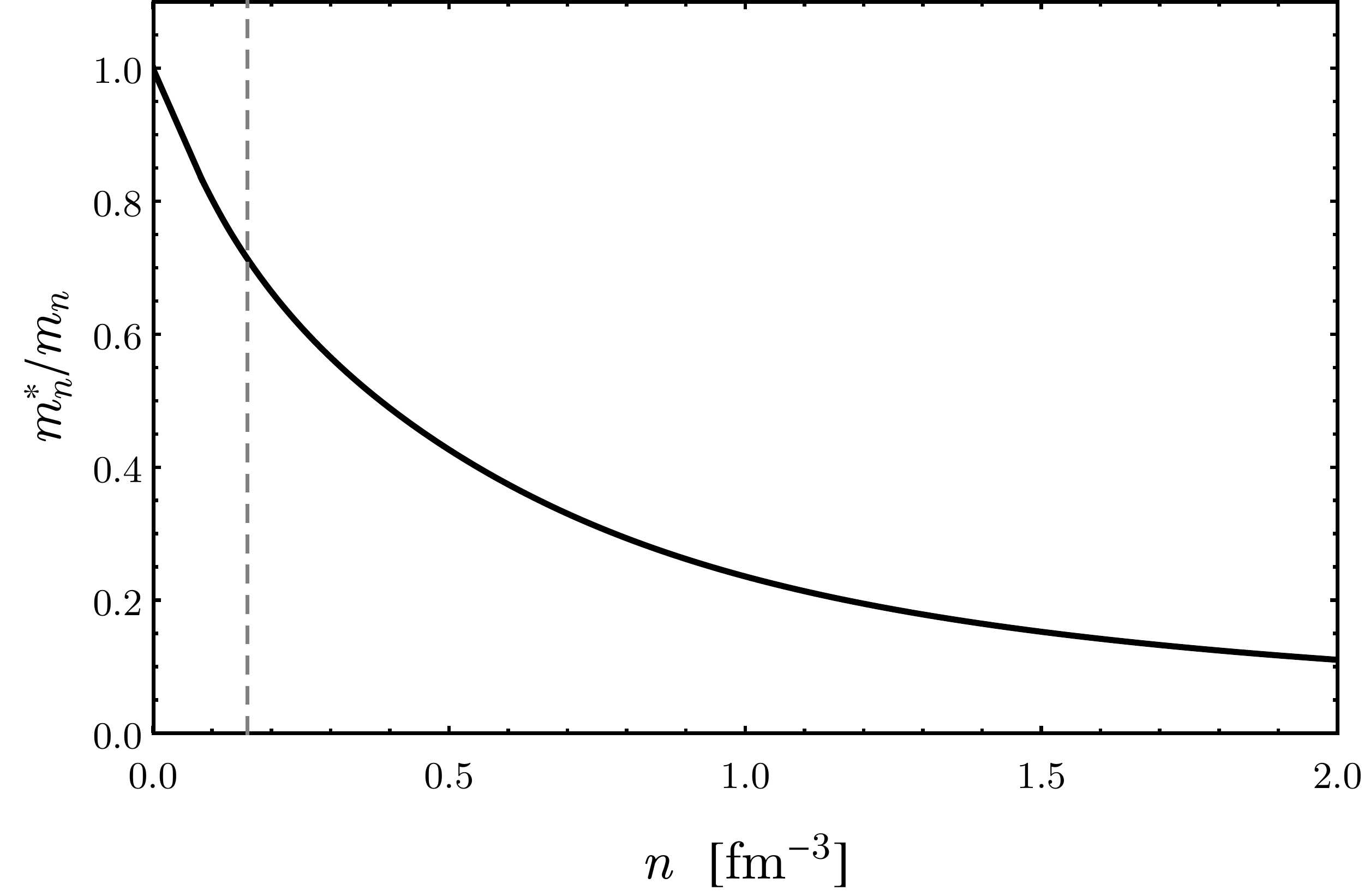} \hfill
 \includegraphics[width = 0.49\textwidth]{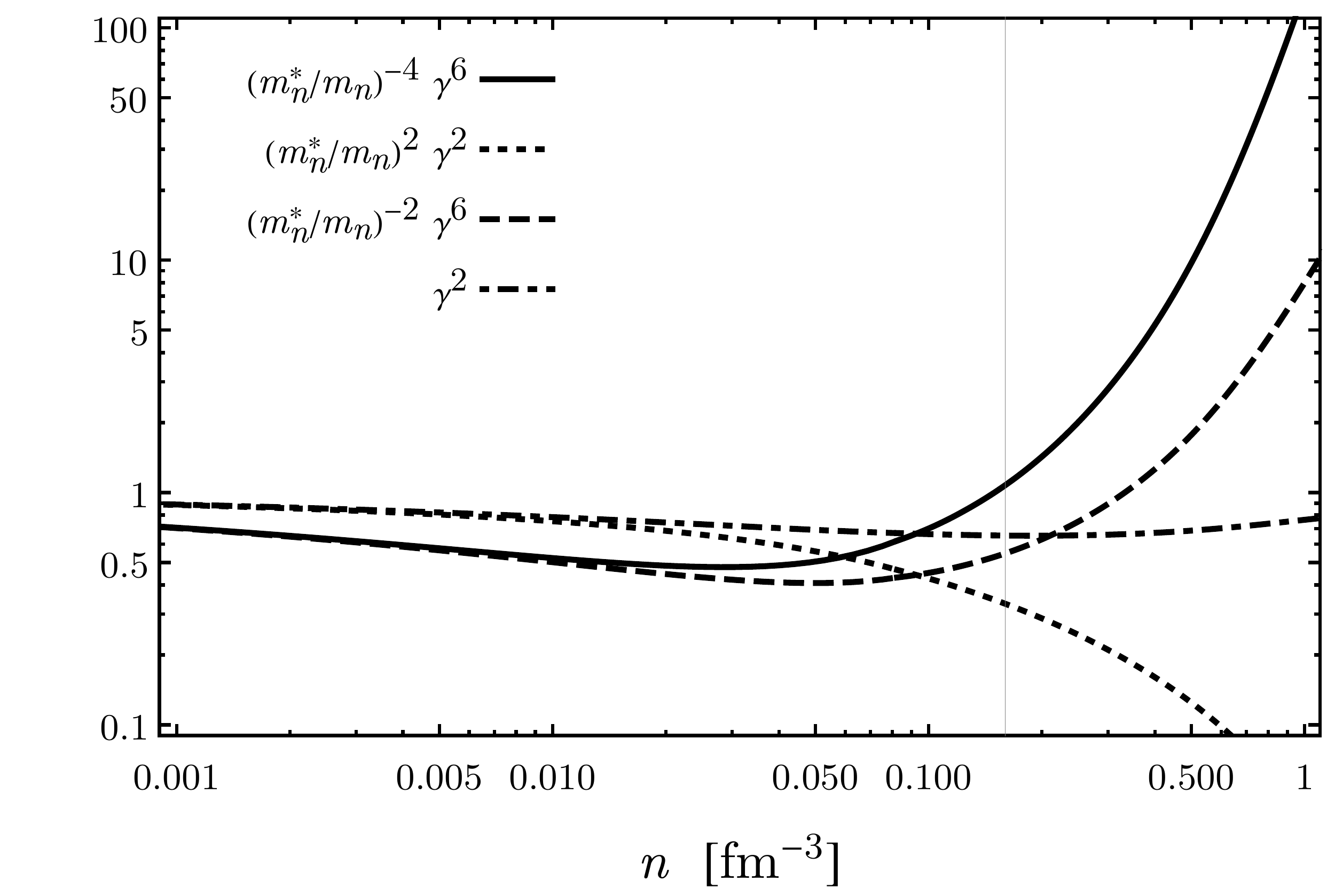}
\caption{\textit{Left:}  The density-dependent neutron effective mass $m_n^\ast$ compared to the vacuum mass $m_n$ for the BSk22 EOS.  \textit{Right:}  The density-dependent correction factors that are added to the axion and neutrino emissivity calculations to account for the medium-dependent effective couplings.  At low density, all of these factors asymptote to $1$.  
}
\label{fig:gamma_factors}
\end{figure*}

For both the axion and neutrino emissivities, we introduced correction factors to account for the effective medium-dependent couplings.  
To assess the effect of these corrections, we plot these factors in Fig.~\ref{fig:gamma_factors} as a function of the neutron number density $n$.  
At the saturation density, $n \simeq 0.16 \ \mathrm{fm}^{-3}$, these factors are all close to $1$ except for $(m_n^\ast/m_n)^2 \gamma^2 \simeq 0.3$, which suppresses the axion PBF emissivities.  
For typical stars in our cooling simulations, the maximal core density can be as large as $n \approx 0.4 \ \mathrm{fm}^{-3}$.  
At these higher densities, $(m_n^\ast/m_n)^2 \gamma^2 \simeq 0.2$ suppresses the axion PBF emissivities, while $(m_n^\ast/m_n)^{-4} \gamma^6 \simeq 5$ enhances the MURCA and neutrino bremsstrahlung rates.  
The factor $\gamma^2 \approx 1$ across the whole range of NS densities, implying a negligible effect on the neutrino PBF emissivities.  
These corrections due to the effective medium-dependent couplings where not taken into account in previous studies of axion limits from NS cooling or SN 1987A.  Thus, as mentioned in the main Letter, we estimate that the SN 1987A suggested upper limits in Ref.~\cite{Carenza:2019pxu} should be a factor $\sim$1.3--1.6 weaker, depending on the EOS, after the medium-dependent couplings have been accounted for.  This is because that work computed the upper limits on axion couplings by requiring $L_a^\infty / L_\nu^\infty < 1$.

\section{Statistical methodology}
\label{app:stat}

Because our analysis includes a large number of nuisance parameters, many of which have degeneracies with each other and the signal parameter $m_a$, upper limits and possible detection significances cannot be interpreted in the asymptotic limit through Wilks' theorem. Instead, we determine limits and significances through MC procedures  (see,~{\it e.g.},~\cite{Cowan:2010js}). We describe those procedures in detail in this section. 

\subsection{Detection Significance}
\label{sec:sig}
We use the test statistic $q_0$ for the discovery of a  signal in order to assess the significance of the axion hypothesis over the null hypothesis. This test statistic is defined as 
\begin{equation}
    q_0 = -2 \ln \frac{\mathcal{L}({\bm d}| m_a = 0, \hat{\bm \theta}(0))}{\mathcal{L}({\bm d}| \hat m_a, \hat{\bm \theta}(\hat m_a ))},
\end{equation}
where $\hat m_a$ is the axion mass at which $\mathcal{L}(\bm d)$ is globally maximized and $\hat{\bm \theta}(m_a)$ is the nuisance parameter vector that maximizes $\mathcal{L}(\bm d)$ at fixed $m_a$.  Recall that $m_a<0$ is allowed, with the axion emissivities multiplied by ${\rm sign}(m_a)$.  In order to assess for mismodeling we perform a two-sided significance test for the axion model, even though only positive axion masses are physical. A $p$-value for the improvement of the goodness-of-fit to the observed data with the inclusion of the axion signal parameter $m_a$ can be obtained by
\begin{equation}
\label{eq:disc_p}
    p = \int_{q_{0, \mathrm{obs}}}^{\infty} f(q_{0})dq_0 \,,
\end{equation}
where $f(q_0)$ is the probability density function of the $q_0$ under the null hypothesis. This $p$-value can then be associated with a significance (in terms of number of $\sigma$) by $\sqrt{\Phi^{-1}(1-p)}$, where $\Phi^{-1}$ is the inverse of the cumulative distribution function of the $\chi^2$-distribution. Since we are not in the asymptotic limit, rather than assuming $f$ is the probability density function of the $\chi^2$ distribution, we will determine it through the following MC procedure.

After fitting the null hypothesis to the observed data, we have $\hat{\bm{\theta}}(0)$, which contain a maximum-likelihood age under the null for each star $\bm{t} = \{\hat t^{i}(0)\}$. We also compute the set of maximum-likelihood luminosities  $\bm{L} = \{L(0,  \hat{\bm{\theta}}^{i}(0)) \}$ from $\hat{\bm{\theta}}(0)$. A single MC realization of the data under the null is then constructed by $\bm{d}_\mathrm{MC}^i = \{\bm{L}^i + \delta L^i ,\sigma_{L}^{i},\bm{t}^i + \delta t^i,\sigma_{t}^{i}\}$ where $\delta L^i$ and $\delta t^i$ are variates drawn from a zero-mean Gaussian distribution with standard deviation $\sigma_L^i$ and $\sigma_t^i$, respectively. These quantities represent the measured luminosity and measured age of the NS within the MC realization. Note that infrequently, we may have $\bm{t}^i + \delta t^i < 0$, which we address by flooring the MC measured age at $0$. For a given realization of the MC data, we compute
\begin{equation}
    q^\mathrm{MC}_0 = 
      -2 \ln \frac{\mathcal{L}({\bm d}_\mathrm{MC}| m_a = 0, \hat{\bm \theta}(0))}{\mathcal{L}({\bm d}_\mathrm{MC}| \hat m_a, \hat{\bm \theta}(\hat m_a ))} \,,
\end{equation}
from which we infer $f(q_0)$ and determine a detection significance associated with $q_0$ as calculated from the observed data. This procedure is performed independently for each combination of EOS and superfluidity model.

\begin{figure}[!t]
\includegraphics[width = 0.49\textwidth]{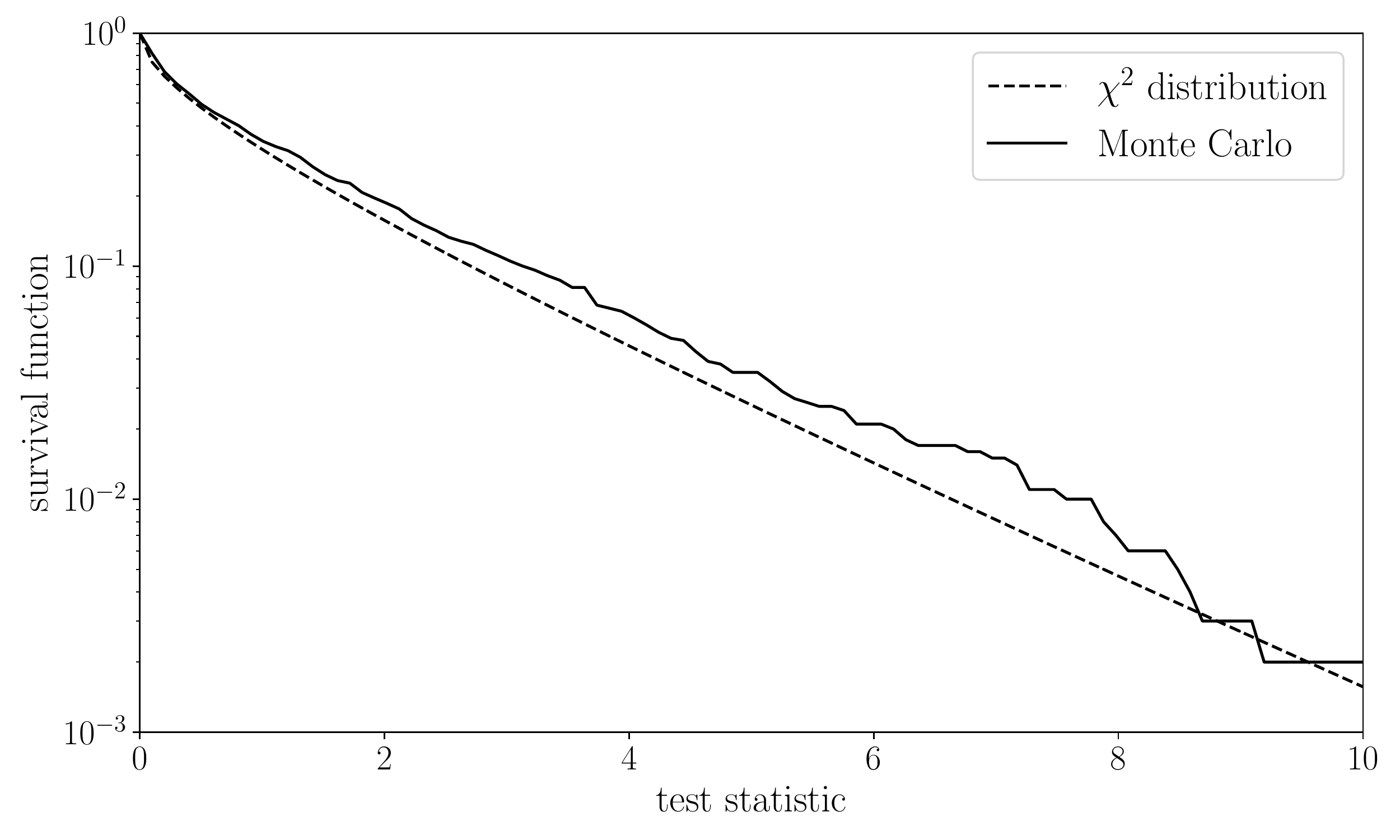}
\includegraphics[width = 0.49\textwidth]{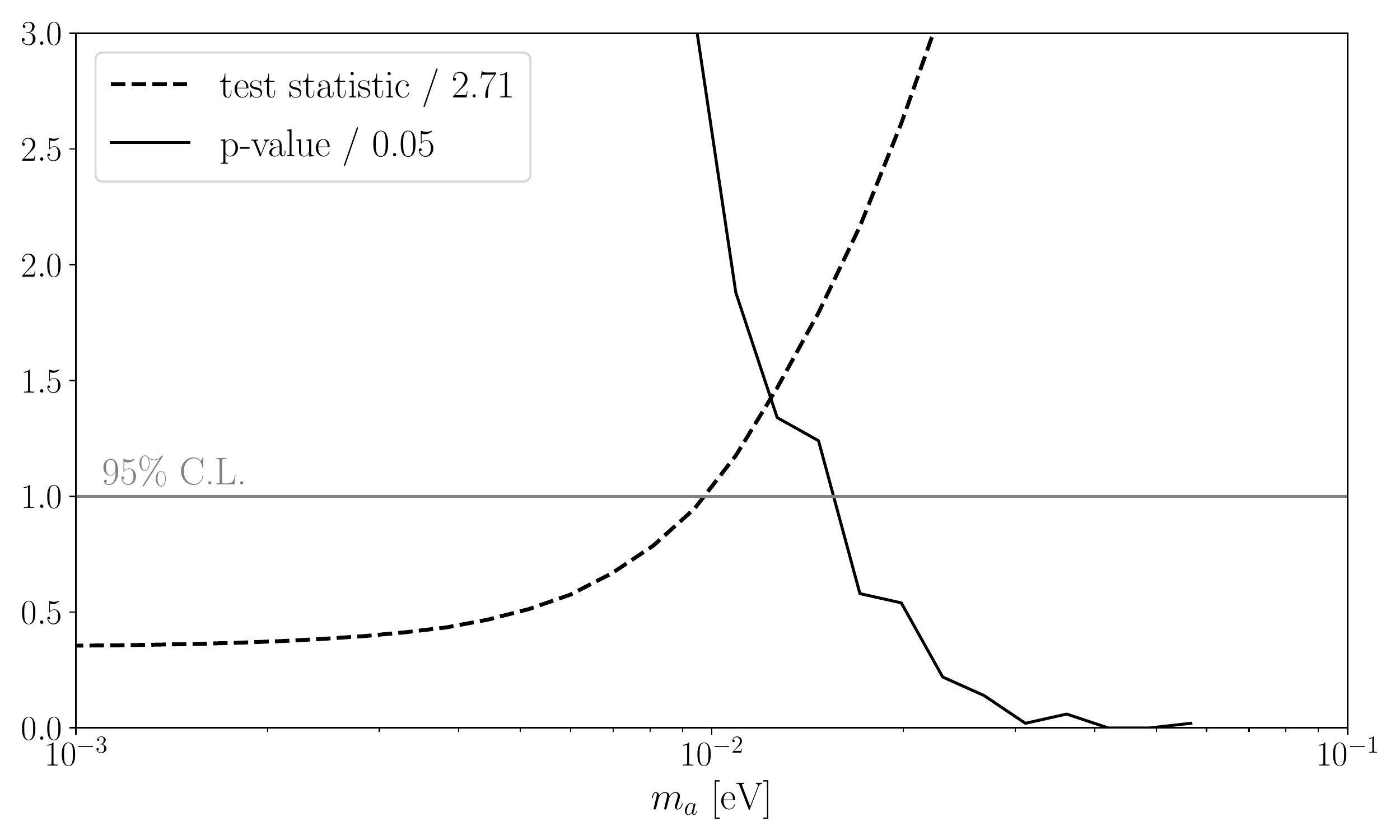}
\caption{The MC distributions used to determine the detection significance of the axion model (left panel) and the 95\% upper limit (right panel) for the KSVZ analysis that leads to the weakest 95\% upper limit (BSk22 EOS and \texttt{SFB-0-0} superfluidity model). We determine the detection significances and 95\% upper limits through MC procedures by repeated simulations of the null and signal hypotheses.  The detection significances are similar to those that would be obtained by assuming Wilks' theorem but the upper limits tend to be more conservative by $\sim$50\% when obtained by MC, as illustrated in the right panel. See text for details.}
\label{fig:pval}
\end{figure}
In the left panel of Fig.~\ref{fig:pval} we illustrate the survival function for the $\chi^2$-distribution (dashed) as a function of the discovery test statistic $q_0$.  We compare this distribution to the distribution of $p$-values, as defined in~\eqref{eq:disc_p}, over an ensemble of $10^3$ MC realizations.  This is the discovery test statistic distribution for our EOS and superfluidity combinations that leads to the weakest upper limit for our KSVZ analysis (BSk22 EOS and \texttt{SFB-0-0} superfluidity model). Note that in this case finite test statistics are somewhat less significant than they would be under the $\chi^2$ distribution, though for other EOS and superfluidity combinations the opposite is true.

\subsection{Upper Limits}
\label{sec:UL}
The procedure for setting a 95\% upper limit follows a similar MC approach as that used to determine a detection significance. We now consider a test statistic $q_{m_a}$ for upper limits defined by 
\begin{equation}
    q_{m_a} = 
    \begin{cases} 
      -2 \ln \frac{\mathcal{L}({\bm d}| m_a, \hat{\bm \theta}(m_a))}{\mathcal{L}({\bm d}| \hat m_a, \hat{\bm \theta}(\hat m_a ))} & \hat m_a < m_a, \\
      0 & \hat m_a \geq m_a, \\
   \end{cases}
   \label{eq:TS_UL}
\end{equation}
The compatibility between the data and a hypothesized value $m_a$ for the axion mass is quantified by the $p$-value
\begin{equation}
    p_{m_a} = \int_{q_{m_a}}^{\infty} f(q_{m_a} | m_a) dq_{_{m_a}} \,,
\end{equation}
where $f(q_{m_a} | m_a)$ is the probability density function for the distribution of $q_{m_a}$ under the assumption that the axion has mass $m_a$. The 95\% upper limit on $m_a$ is then determined at $m_a^{95}$ where $p_{m_a^{95}} = 0.05$.  As in the case of determining the distribution relevant for detection significance, we will use a MC procedure to determine probability density functions, though now we are determining through MC a family of distributions parametrized by the assumed value of $m_a$.

Specifically, for a range of values of $m_a$, we determine $\hat{\bm{\theta}}(m_a)$, providing the maximum-likelihood estimate of the age under the assumed $m_a$ and enabling us to calculate maximum-likelihood luminosities  $\bm{L}(m_a) = \{L(m_a,  \hat{\bm{\theta}}^{i}(0)) \}$ from $\hat{\bm{\theta}}(m_a)$. Similar to before, a single MC realization under the assumed $m_a$ is constructed by $\bm{d}_\mathrm{MC}^i = \{\bm{L}^i + \delta L^i ,\sigma_{L}^{i},\bm{t}^i + \delta t^i,\sigma_{t}^{i}\}$. For each MC realization, we compute $q^\mathrm{MC}_{m_a}$ defined by
\begin{equation}
    q_{m_a}^\mathrm{MC} = 
    \begin{cases} 
      -2 \ln \frac{\mathcal{L}({\bm d_\mathrm{MC}}| m_a, \hat{\bm \theta}(m_a))}{\mathcal{L}({\bm d_\mathrm{MC}}| \hat m_a, \hat{\bm \theta}(\hat m_a ))} & \hat m_a < m_a, \\
      0 & \hat m_a \geq m_a, \\
   \end{cases}
\end{equation}
and then calculate $p_{m_a}$ from the MC distribution. We then vary $m_a$ until $p_{m_a} = 0.05$ to determine our 95\% upper limit. As before, this procedure is performed independently for each combination of EOS and superfluidity model.

In the right panel of Fig.~\ref{fig:pval} we illustrate the $p_{m_a}$ that we determine through the MC procedure for the KSVZ analysis that leads to the weakest limit (BSk22 EOS and \texttt{SBF-0-0} superfluidity model, as in the left panel).  Note that the $p$-value distribution has been rescaled such that the one-sided 95\% upper limit is achieved when the curve crosses unity. On the same figure we illustrate the test statistic itself, as defined in~\eqref{eq:TS_UL}.
In the asymptotic limit where Wilks' theorem holds the 95\% one-sided upper limit should be given by where the  test statistic crosses $\sim$$2.71$ (see, {\it e.g.},~\cite{Cowan:2010js}), though again we have rescaled the test statistic such that the Wilks' limit is achieved for the curve crossing unity. Comparing the limit obtained through the MC procedure to that obtained by assuming Wilks' theorem we see that the MC limit is more conservative by $\sim$50\% in $m_a$. 

\section{Extended Results}
\label{app:extended}

\begin{figure}[!htb]
\includegraphics[width = 0.7\textwidth]{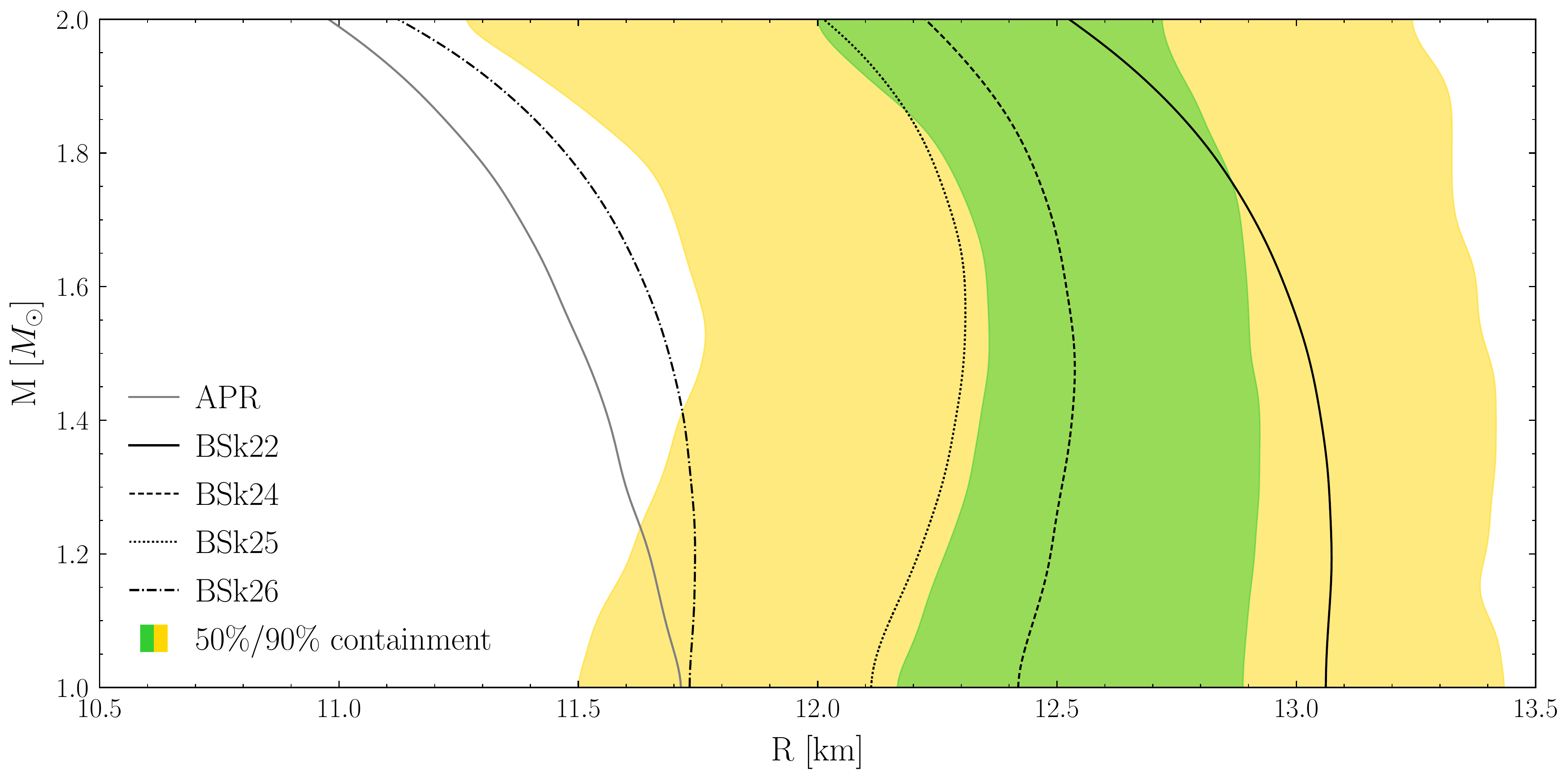}
\caption{
Green and gold bands show the containment regions (at indicated confidence) for the NS mass and radius as constructed in~\cite{Miller:2021qha}.  That work made use of simultaneous mass-radius measurements of two NSs, PSR J0030~\cite{Riley:2019yda} and PSR J0740~\cite{Miller:2021qha}, with {\it NICER} data, in conjunction with gravitational wave data from NS mergers.  On top of the containment regions we illustrate the mass-radius predictions from the five EOS considered in this work.  The APR and BSk26 EOS are not consistent within 90\% with the mass-radius data and are thus not considered in our fiducial analyses, though results with these EOS are presented in the SM. }
\label{fig:EOS_MR}
\end{figure}

In this section we present additional results related to the analyses discussed in the main Letter.   
The EOS that are consistent with the mass-radius relation (BSk22, BSk24, BSk25) determined by~\cite{Miller:2021qha} and inconsistent (APR, BSk26) are illustrated in Fig.~\ref{fig:EOS_MR}.  
Note that the green and gold bands show the containment regions determined from that work at the indicated confidence.  
Ref.~\cite{Miller:2021qha} combined mass-radius measurements of PSR J0030~\cite{Riley:2019yda} and PSR J0740~\cite{Miller:2021qha}, made with {\it NICER} data, with gravitational wave data from NS mergers in the context of a non-parametric mass-radius model based off of Gaussian Process modeling.

\subsection{Extended results for fiducial analyses}

In Tab.~\ref{tab:upper} we present the full results from the analyses of the different EOS and superfluidity combinations for the KSVZ axion.  We present the 95\% upper limits ($m_a^{95}$), the best-fit axion masses ($\hat m_a$), allowing the best-fit masses to be negative with the axion luminosities multiplied by ${\rm sign}(m_a)$, the significance of the axion model over the null hypothesis under the two-sided test, and the absolute $\chi^2$ value of the null hypothesis test. The significance is quoted in terms of $\sigma$ as determined by our MC procedure described in Sec.~\ref{sec:sig}. The $\chi^2$ value of the null hypothesis is defined by
\es{eq:chi2}{
\chi^2 \equiv \sum_{i} {\left[ L(0,{\bm{\hat \theta}}^i) - L_0^i \right]^2 \over (\sigma_L^i)^2} + {({\hat t}^i - t_0^i)^2 \over (\sigma_t^i)^2} \,,
}
with the sum over the five NSs and the notation as in~\eqref{eq:indL}.  Note that ${\bm{\hat \theta}}^i$ denotes the best-fit model parameter vector under the null hypothesis ($m_a = 0$).  Smaller values of $\chi^2$ denote better fits of the null hypothesis, though keep in mind that many of the $\chi^2$ values are smaller than unity because of the large number of nuisance parameters.  Still, the $\chi^2$ values clearly show that some superfluidity and EOS models provide worse fits to the data than others.  The best-fitting model under the null hypothesis is the BSk22 EOS with no superfluidity.
In Tabs.~\ref{tab:1856_nuis}, \ref{tab:1308_nuis}, \ref{tab:0720_nuis}, \ref{tab:1605_nuis}, and \ref{tab:0659_nuis}, we show the best-fit nuisance parameters for the individual NSs under the null hypothesis for the different EOS and superfluidity combinations.  Note that many of the NSs have best-fit masses near or at $1 M_\odot$, which is the lower edge of our mass prior. However, as we show, the dependence of our results on the NS mass is relatively minor, so long as the mass is not large enough for the direct URCA process to be important.

\setlength{\tabcolsep}{6pt}
\begin{table}
\centering
\begin{tabular}{c | c c c c c} 
 
 & BSk22 & BSk24 & BSk25 & Bsk26 & APR \\ 
 \hline 
 & $m_a^{95}$=14 meV & $m_a^{95}$=12 meV & $m_a^{95}$=8.8 meV & $m_a^{95}$=13 meV & $m_a^{95}$=4.7 meV\\
0-0-0 & $\hat m_a$=-8.1 meV & $\hat m_a$=-9.4 meV & $\hat m_a$=-9.4 meV & $\hat m_a$=-11 meV & $\hat m_a$=-11 meV\\
 & $\sigma=$0.36 & $\sigma=$0.8 & $\sigma=$1.1 & $\sigma=$0.83 & $\sigma=$1.4\\
 & $\chi^2=$0.46 & $\chi^2=$0.99 & $\chi^2=$1.8 & $\chi^2=$1 & $\chi^2=$3\\
 \hline 
 & $m_a^{95}$=16 meV & $m_a^{95}$=13 meV & $m_a^{95}$=9.3 meV & $m_a^{95}$=14 meV & $m_a^{95}$=3.4 meV\\
SFB-0-0 & $\hat m_a$=-9.4 meV & $\hat m_a$=-9.4 meV & $\hat m_a$=-11 meV & $\hat m_a$=-13 meV & $\hat m_a$=-13 meV\\
 & $\sigma=$0.92 & $\sigma=$1.1 & $\sigma=$1.5 & $\sigma=$1.2 & $\sigma=$1.9\\
 & $\chi^2=$1.2 & $\chi^2=$1.9 & $\chi^2=$2.9 & $\chi^2=$2 & $\chi^2=$4.8\\
 \hline 
 & $m_a^{95}$=15 meV & $m_a^{95}$=10 meV & $m_a^{95}$=6 meV & $m_a^{95}$=10 meV & $m_a^{95}$=4.6 meV\\
SFB-0-T73 & $\hat m_a$=-8.1 meV & $\hat m_a$=-9.4 meV & $\hat m_a$=-11 meV & $\hat m_a$=-11 meV & $\hat m_a$=-13 meV\\
 & $\sigma=$0.83 & $\sigma=$1.5 & $\sigma=$1.6 & $\sigma=$1.3 & $\sigma=$1.9\\
 & $\chi^2=$1.8 & $\chi^2=$2.9 & $\chi^2=$4.2 & $\chi^2=$2.3 & $\chi^2=$5.2\\
 \hline 

\end{tabular}
\caption{
 \label{tab:upper}
95\% C.L. limit $m_a^{95}$, best-fit axion mass $\hat m_a$,
and significance $\sigma$ of the best fit under the KSVZ axion model for different 
combinations of EOS and superfluidity model. We also provide the $\chi^2$ quantity which describes the goodness-of-fit of the null model. Note that significance $\sigma$ is computed through the MC procedure described in Sec.~\ref{sec:sig} but is presented in the equivalent number of $\sigma$ for a $\chi^2$-distributed discovery test statistic.}
\end{table}

\setlength{\tabcolsep}{6pt}
\begin{table}
\centering
\begin{tabular}{c | c c c c c} 
 
\bf{J1856} & BSk22 & BSk24 & BSk25 & Bsk26 & APR \\ 
 \hline 
 & $M_\text{NS}=1.0 M_\odot$  & $M_\text{NS}=1.6 M_\odot$  & $M_\text{NS}=1.4 M_\odot$  & $M_\text{NS}=1.4 M_\odot$  & $M_\text{NS}=1.0 M_\odot$ \\
0-0-0 & $\Delta M=10^{-6}$  & $\Delta M=10^{-10}$  & $\Delta M=10^{-10}$  & $\Delta M=10^{-10}$  & $\Delta M=10^{-10}$ \\
 \hline 
 & $M_\text{NS}=1.0 M_\odot$  & $M_\text{NS}=1.4 M_\odot$  & $M_\text{NS}=1.0 M_\odot$  & $M_\text{NS}=1.2 M_\odot$  & $M_\text{NS}=1.8 M_\odot$ \\
SFB-0-0 & $\Delta M=10^{-8}$  & $\Delta M=10^{-10}$  & $\Delta M=10^{-10}$  & $\Delta M=10^{-10}$  & $\Delta M=10^{-12}$ \\
 \hline 
 & $M_\text{NS}=1.0 M_\odot$  & $M_\text{NS}=1.0 M_\odot$  & $M_\text{NS}=1.0 M_\odot$  & $M_\text{NS}=1.0 M_\odot$  & $M_\text{NS}=1.8 M_\odot$ \\
SFB-0-T73 & $\Delta M=10^{-10}$  & $\Delta M=10^{-10}$  & $\Delta M=10^{-10}$  & $\Delta M=10^{-10}$  & $\Delta M=10^{-12}$ \\
 \hline 

\end{tabular}
\caption{
\label{tab:1856_nuis}
Best-fit nuisance parameters assuming no axion for the NS J1856
for different combinations of EOS and superfluidity model.}
\end{table}

\setlength{\tabcolsep}{6pt}
\begin{table}
\centering
\begin{tabular}{c | c c c c c} 
 
\bf{J1308} & BSk22 & BSk24 & BSk25 & Bsk26 & APR \\ 
 \hline 
 & $M_\text{NS}=1.0 M_\odot$  & $M_\text{NS}=1.0 M_\odot$  & $M_\text{NS}=1.0 M_\odot$  & $M_\text{NS}=1.0 M_\odot$  & $M_\text{NS}=1.0 M_\odot$ \\
0-0-0 & $\Delta M=10^{-12}$  & $\Delta M=10^{-12}$  & $\Delta M=10^{-12}$  & $\Delta M=10^{-12}$  & $\Delta M=10^{-12}$ \\
 \hline 
 & $M_\text{NS}=1.0 M_\odot$  & $M_\text{NS}=1.2 M_\odot$  & $M_\text{NS}=1.2 M_\odot$  & $M_\text{NS}=1.0 M_\odot$  & $M_\text{NS}=1.6 M_\odot$ \\
SFB-0-0 & $\Delta M=10^{-12}$  & $\Delta M=10^{-12}$  & $\Delta M=10^{-12}$  & $\Delta M=10^{-12}$  & $\Delta M=10^{-14}$ \\
 \hline 
 & $M_\text{NS}=1.0 M_\odot$  & $M_\text{NS}=1.0 M_\odot$  & $M_\text{NS}=1.0 M_\odot$  & $M_\text{NS}=1.0 M_\odot$  & $M_\text{NS}=1.6 M_\odot$ \\
SFB-0-T73 & $\Delta M=10^{-12}$  & $\Delta M=10^{-12}$  & $\Delta M=10^{-12}$  & $\Delta M=10^{-12}$  & $\Delta M=10^{-14}$ \\
 \hline 

\end{tabular}
\caption{
\label{tab:1308_nuis}
Best-fit nuisance parameters assuming no axion for the NS J1308
for different combinations of EOS and superfluidity model.}
\end{table}

\setlength{\tabcolsep}{6pt}
\begin{table}
\centering
\begin{tabular}{c | c c c c c} 
 
\bf{J0720} & BSk22 & BSk24 & BSk25 & Bsk26 & APR \\ 
 \hline 
 & $M_\text{NS}=1.0 M_\odot$  & $M_\text{NS}=1.0 M_\odot$  & $M_\text{NS}=1.0 M_\odot$  & $M_\text{NS}=1.0 M_\odot$  & $M_\text{NS}=1.0 M_\odot$ \\
0-0-0 & $\Delta M=10^{-14}$  & $\Delta M=10^{-14}$  & $\Delta M=10^{-14}$  & $\Delta M=10^{-14}$  & $\Delta M=10^{-16}$ \\
 \hline 
 & $M_\text{NS}=1.0 M_\odot$  & $M_\text{NS}=1.2 M_\odot$  & $M_\text{NS}=1.2 M_\odot$  & $M_\text{NS}=1.2 M_\odot$  & $M_\text{NS}=1.2 M_\odot$ \\
SFB-0-0 & $\Delta M=10^{-14}$  & $\Delta M=10^{-16}$  & $\Delta M=10^{-16}$  & $\Delta M=10^{-16}$  & $\Delta M=10^{-16}$ \\
 \hline 
 & $M_\text{NS}=1.0 M_\odot$  & $M_\text{NS}=1.2 M_\odot$  & $M_\text{NS}=1.0 M_\odot$  & $M_\text{NS}=1.0 M_\odot$  & $M_\text{NS}=1.4 M_\odot$ \\
SFB-0-T73 & $\Delta M=10^{-14}$  & $\Delta M=10^{-16}$  & $\Delta M=10^{-18}$  & $\Delta M=10^{-16}$  & $\Delta M=10^{-20}$ \\
 \hline 

\end{tabular}
\caption{
\label{tab:0720_nuis}
Best-fit nuisance parameters assuming no axion for the NS J0720
for different combinations of EOS and superfluidity model.}
\end{table}

\setlength{\tabcolsep}{6pt}
\begin{table}
\centering
\begin{tabular}{c | c c c c c} 
 
\bf{J1605} & BSk22 & BSk24 & BSk25 & Bsk26 & APR \\ 
 \hline 
 & $M_\text{NS}=1.0 M_\odot$  & $M_\text{NS}=1.2 M_\odot$  & $M_\text{NS}=1.0 M_\odot$  & $M_\text{NS}=1.0 M_\odot$  & $M_\text{NS}=1.0 M_\odot$ \\
0-0-0 & $\Delta M=10^{-12}$  & $\Delta M=10^{-12}$  & $\Delta M=10^{-12}$  & $\Delta M=10^{-12}$  & $\Delta M=10^{-12}$ \\
 \hline 
 & $M_\text{NS}=1.0 M_\odot$  & $M_\text{NS}=1.2 M_\odot$  & $M_\text{NS}=1.2 M_\odot$  & $M_\text{NS}=1.0 M_\odot$  & $M_\text{NS}=1.4 M_\odot$ \\
SFB-0-0 & $\Delta M=10^{-12}$  & $\Delta M=10^{-12}$  & $\Delta M=10^{-12}$  & $\Delta M=10^{-12}$  & $\Delta M=10^{-14}$ \\
 \hline 
 & $M_\text{NS}=1.0 M_\odot$  & $M_\text{NS}=1.0 M_\odot$  & $M_\text{NS}=1.0 M_\odot$  & $M_\text{NS}=1.0 M_\odot$  & $M_\text{NS}=1.4 M_\odot$ \\
SFB-0-T73 & $\Delta M=10^{-12}$  & $\Delta M=10^{-12}$  & $\Delta M=10^{-12}$  & $\Delta M=10^{-12}$  & $\Delta M=10^{-14}$ \\
 \hline 

\end{tabular}
\caption{
\label{tab:1605_nuis}
Best-fit nuisance parameters assuming no axion for the NS J1605
for different combinations of EOS and superfluidity model.}
\end{table}

\setlength{\tabcolsep}{6pt}
\begin{table}
\centering
\begin{tabular}{c | c c c c c} 
 
\bf{J0659} & BSk22 & BSk24 & BSk25 & Bsk26 & APR \\ 
 \hline 
 & $M_\text{NS}=1.0 M_\odot$  & $M_\text{NS}=1.2 M_\odot$  & $M_\text{NS}=1.0 M_\odot$  & $M_\text{NS}=1.0 M_\odot$  & $M_\text{NS}=1.8 M_\odot$ \\
0-0-0 & $\Delta M=10^{-8}$  & $\Delta M=10^{-10}$  & $\Delta M=10^{-10}$  & $\Delta M=10^{-10}$  & $\Delta M=10^{-12}$ \\
 \hline 
 & $M_\text{NS}=1.0 M_\odot$  & $M_\text{NS}=1.0 M_\odot$  & $M_\text{NS}=1.0 M_\odot$  & $M_\text{NS}=2.0 M_\odot$  & $M_\text{NS}=1.6 M_\odot$ \\
SFB-0-0 & $\Delta M=10^{-10}$  & $\Delta M=10^{-10}$  & $\Delta M=10^{-20}$  & $\Delta M=10^{-20}$  & $\Delta M=10^{-12}$ \\
 \hline 
 & $M_\text{NS}=1.0 M_\odot$  & $M_\text{NS}=1.0 M_\odot$  & $M_\text{NS}=1.6 M_\odot$  & $M_\text{NS}=1.8 M_\odot$  & $M_\text{NS}=1.6 M_\odot$ \\
SFB-0-T73 & $\Delta M=10^{-10}$  & $\Delta M=10^{-10}$  & $\Delta M=10^{-12}$  & $\Delta M=10^{-12}$  & $\Delta M=10^{-12}$ \\
 \hline 

\end{tabular}
\caption{
\label{tab:0659_nuis}
Best-fit nuisance parameters assuming no axion for the NS J0659
for different combinations of EOS and superfluidity model.}
\end{table}

\begin{figure}[!htb]
\includegraphics[width = 0.7\textwidth]{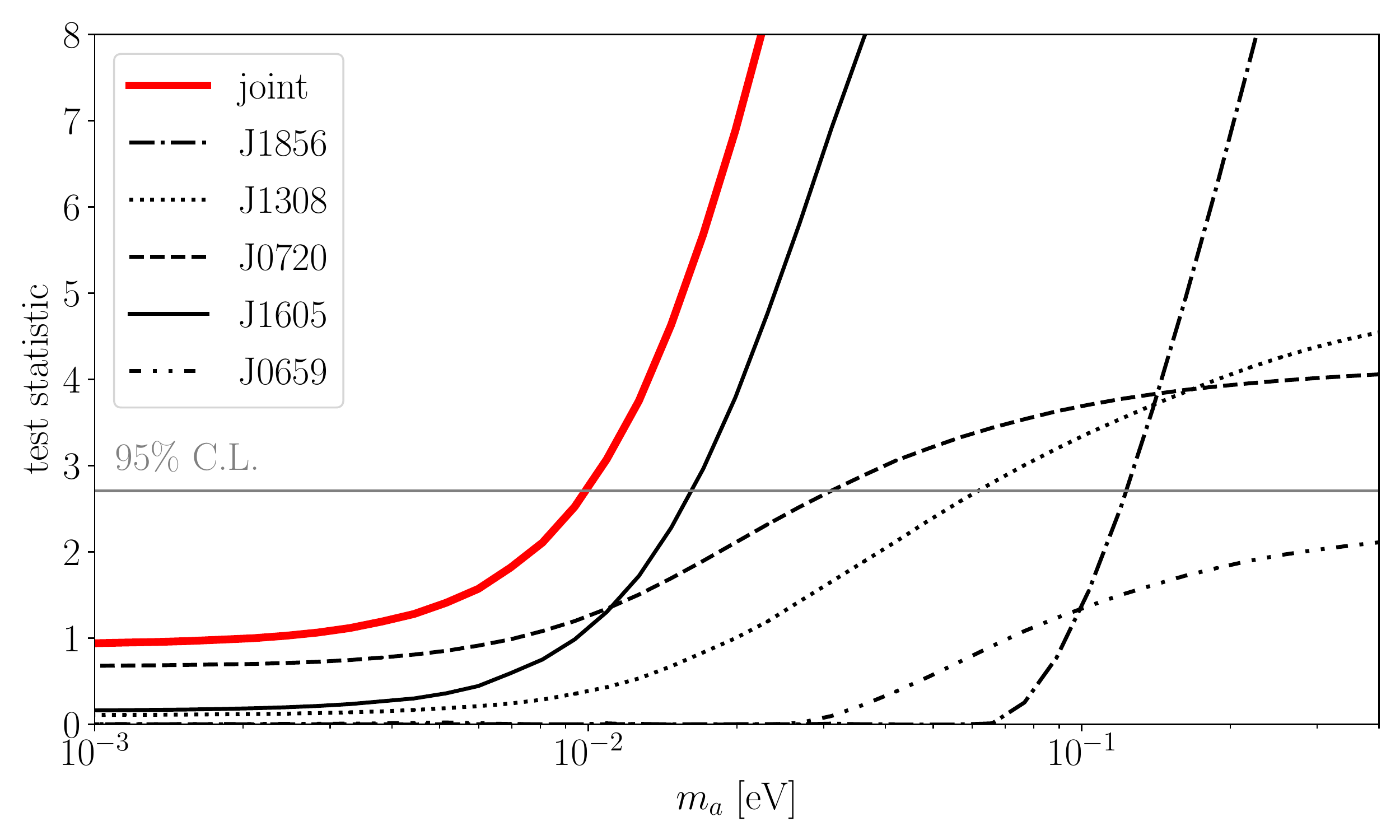}
\caption{The test statistic for upper limits, defined in~\eqref{eq:TS_UL} and in the context of the KSVZ axion model, for the individual NSs.  These curves assume the \texttt{SFB-0-0} superfluidity model and the BSk22 EOS.  Assuming Wilks' theorem, the 95\% upper limit is given by where the test statistic is equal to $\sim$2.71, as indicated.  The most constraining NS is J1605.}
\label{fig:tsinv}
\end{figure}

In Fig.~\ref{fig:tsinv} we show the test statistic for upper limits, defined in~\eqref{eq:TS_UL}, for the individual NSs considered in this work as functions of the KSVZ axion mass $m_a$.  Note that these curves extend to negative masses, though they are only illustrated for positive masses.  We illustrate the test statistics assuming the \texttt{SFB-0-0} superfluidity model and the BSk22 EOS, since that leads to the most conservative limits for the KSVZ axion.  As described in Sec.~\ref{sec:UL}, assuming Wilks' theorem is not a good approximation in determining the upper limit and leads to an overestimate of the limit by $\sim$50\%.  Still, for the purpose of comparing the relative importance of different NSs it is instructive to compare their upper-limit test statistics.  Recall that assuming Wilks' theorem the 95\% upper limits are determined by where these curves cross $\sim$2.71.  From Fig.~\ref{fig:tsinv} we see that the most constraining NS is J1605, followed by J0720 and J1308. 

\begin{figure}[!htb]
\includegraphics[width = 0.6\textwidth]{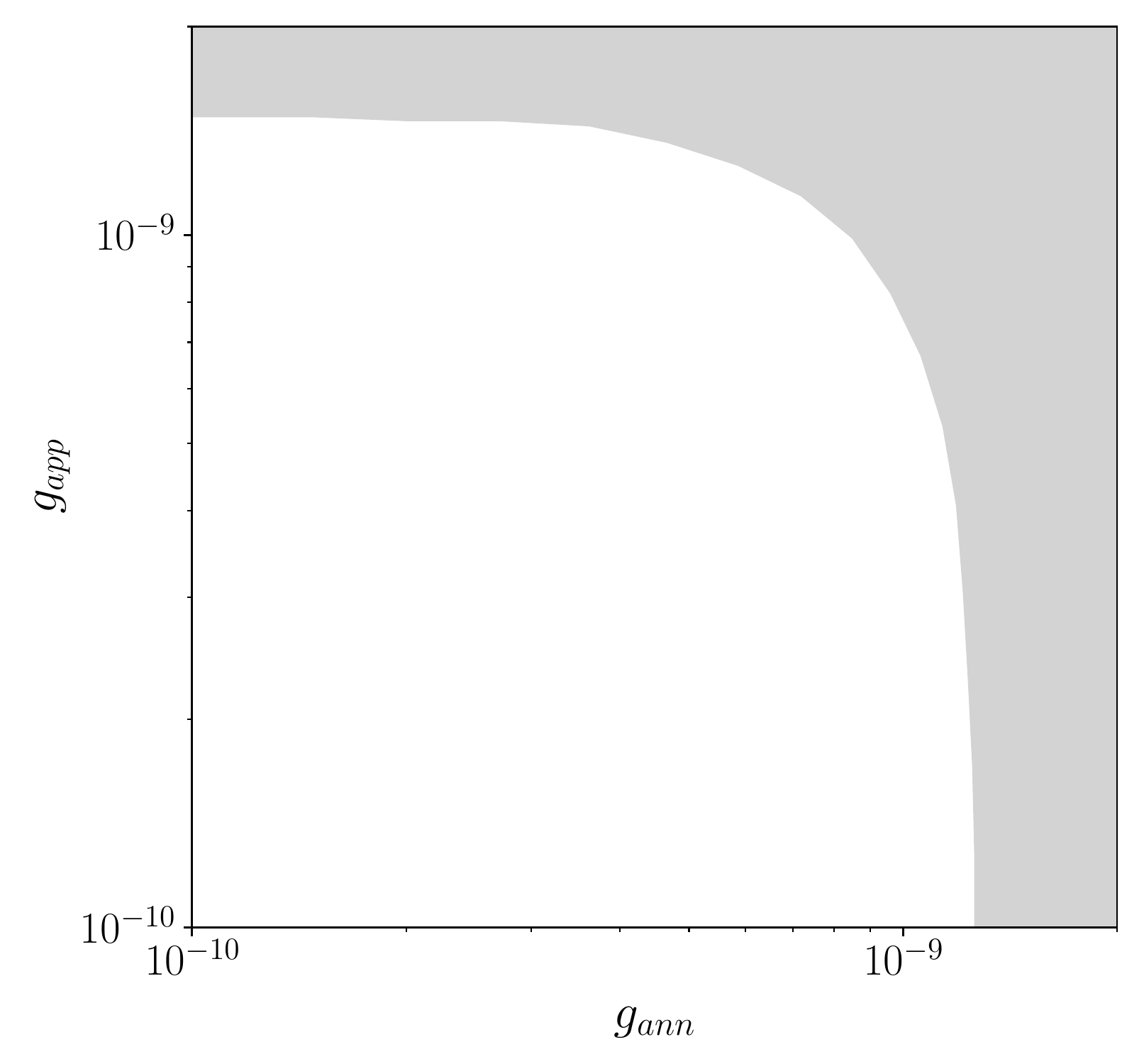}
\caption{{The one-sided 95\% upper limits from this work in the plane of axion-neutron ($g_{ann}$) and axion-proton ($g_{app}$) couplings.  The shaded region is excluded by our analysis; see text for details. } }
\label{fig:ganngapp}
\end{figure}

{In the main Letter we interpreted our results in the context of the KSVZ and DFSZ QCD axion models.  In Fig.~\ref{fig:ganngapp} we take a more phenomenological approach and consider axion models characterized by coupling strengths $g_{ann}$ and $g_{app}$ to neutrons and protons, respectively.  The shaded region in that parameter space is excluded by our analysis at 95\% confidence.  To construct this figure we fix the ratio $g_{ann} / g_{app}$ and then construct the likelihood profile as a function of $g_{ann}$.  The presented limits are then the one-sided 95\% upper limits constructed from our MC procedure.}

\subsection{Axions in younger neutron stars}

\begin{table}[t]
\begin{tabular}{|c|c|c|c|c|}
\hline
Name   & $L_\gamma^\infty$ [$10^{33}$ erg/s] & Age [yr] & Refs                                                                                            \\ \hline
Cas A             & \textemdash         & $330 \pm 19$ & \cite{1980JHA....11....1A,Fesen:2006zma} \\ \hline
J173203           & $17 \pm 10$          & $(4 \pm 2) \times 10^3$ &  \cite{Klochkov:2014ola,Acero:2015caa,Cui:2016pzm,Maxted:2017xfs}                          \\ \hline
J172054           & $11.4 \pm 4.4$           & $650 \pm 50$ & \cite{Ho:2021hwy,2011ApJ...731...70L,Mayer:2021wvs}                             \\ \hline
J0357           & $0.015 \pm 0.011$          & $(7.5\pm5.5)\times10^5$ &  \cite{Kirichenko:2014ona}       \\ \hline
J0538           & $1.09 \pm 0.47$          & $(1.10\pm0.3)\times10^5$ &  \cite{Ng:2006vh,Suzuki:2021ium}       \\ \hline
\end{tabular}
\caption{\label{tab:supp-NSs} The properties of the supplemental NSs considered in this work: CXOU J232327.8+584842 (Cas A), XMMS J173203-344518 (J173203), XMMU J172054.5-372652 (J172054), PSR J0357+3205 (J0357), PSR J0538+2817 (J0538).   Note that for Cas A we use measurements of its temperature $T = (200.0 \pm 5) \times 10^4$ K (with the uncertainty roughly accounting for systematic differences in $T$ measurements between different analyses, in addition to statistical uncertainties) and its temperature derivative $\dot T / T = 0.0011 \pm 0.00046$ yr$^{-1}$ (with the value taken as that with fixed hydrogen column density, since it is the most conservative) rather than the luminosity measurement~\cite{Posselt:2018xaf}.  Also, to match the conventions of~\cite{Posselt:2018xaf}, the age of Cas A is presented at the reference epoch of 2011.49. }
\end{table}

\begin{figure}[!htb]
\includegraphics[width = 0.7\textwidth]{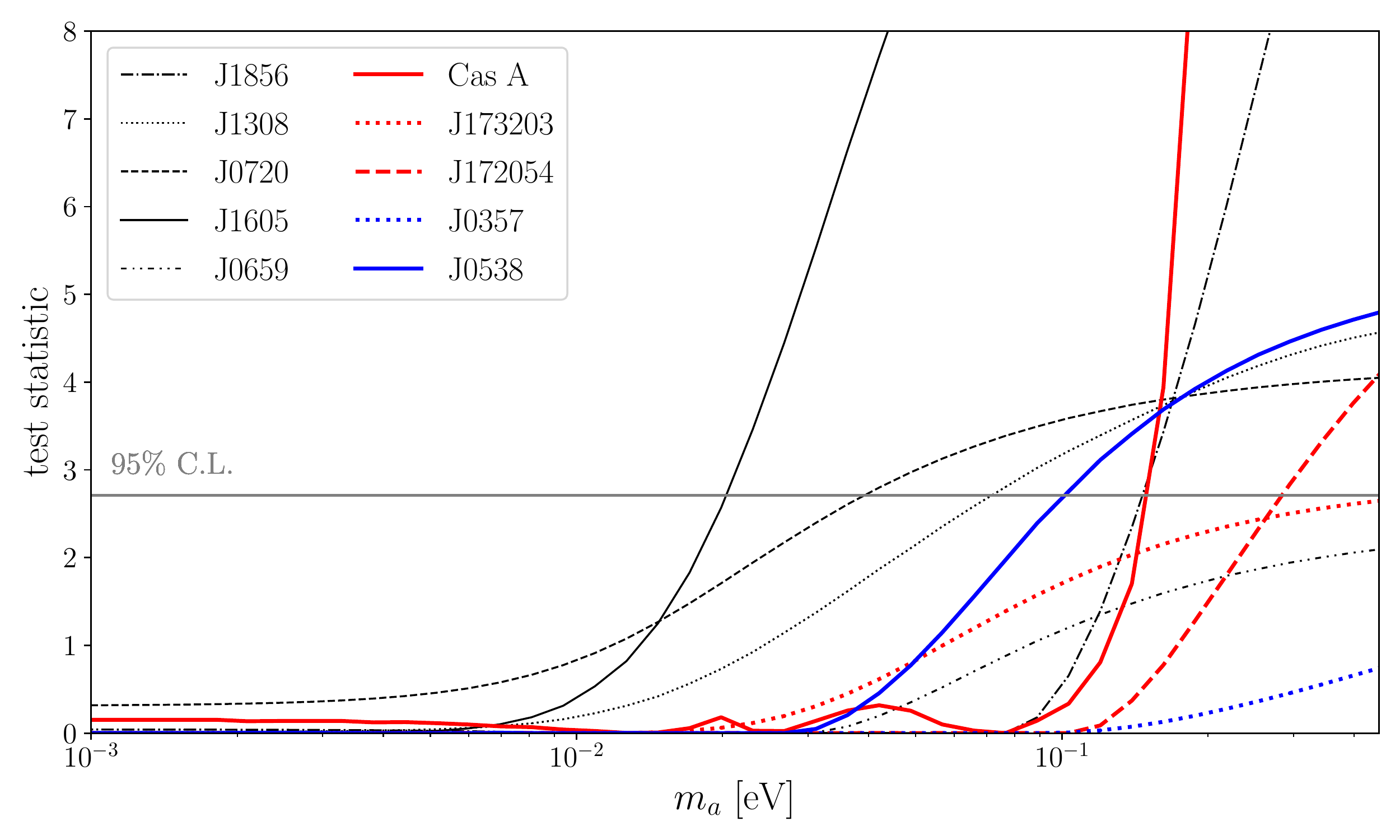}
\caption{As in Fig.~\ref{fig:tsinv} but including the five younger NSs we consider in this work, whose properties are given in Tab.~\ref{tab:supp-NSs}.  Note that unlike in Fig.~\ref{fig:tsinv} here we display the test statistics for the EOS and superfluidity combinations that lead to the weakest limits for the individual NSs not the combination that leads to the weakest limit in a joint analysis.  }
\label{fig:tsinv_CA}
\end{figure}

In this work we focus on older NSs, with ages over $\sim$10$^5$ yrs.  We find that younger NSs, including Cas A, are typically less constraining in the context of our modeling procedure, {though a few younger NSs have comparable sensitivity.} This point is illustrated in Fig.~\ref{fig:tsinv_CA}, where we compare the upper-limit test statistic (for the KSVZ axion) from the individual NSs considered in this work with those we determine from analyses of five younger NSs.  Note that in Fig.~\ref{fig:tsinv_CA}  we show the EOS and superfluidity combinations that lead to the weakest limits, as estimated by the test statistic assuming Wilks' theorem, for the individual NSs.  This is unlike in Fig.~\ref{fig:tsinv} where we show the individual test statistics for the EOS and superfluidity model that gives the weakest limit in the joint analysis over all five old, isolated NSs. The properties of the additional NSs are given in Tab.~\ref{tab:supp-NSs}.  For Cas A we also include the temperature derivative measurement from~\cite{Posselt:2018xaf}, who measure $\dot T / T = (0.011 \pm 0.0046) / (10 \, \, {\rm yrs})$ assuming the Hydrogen column density was constant over all {\it Chandra} observations of Cas A.  As mentioned in the main body, previous to~\cite{Posselt:2018xaf} analyses of Cas A (such as~\cite{2010ApJ...719L.167H,Elshamouty:2013nfa}) assumed strong superfluidity in order to explain the rapidly changing $\dot T / T$, but~\cite{Posselt:2018xaf} pointed out that much of this cooling was instrumental in nature.  Indeed, as shown in~\ref{fig:tsinv_CA}, the Cas A cooling data are consistent with the null hypothesis without $^3P_2$ superfluidity, since the Cas A curve in Fig.~\ref{fig:tsinv_CA} is for the model with no superfluidity and the BSk22 EOS. 
J173203 has also been the target of previous axion searches~\cite{Beznogov:2018fda,Leinson:2019cqv} that found $m_{a,{\rm KSVZ}} < 0.085$ eV at 90\% confidence. These works assumed that the age of J173203 was $\sim$27 kyr~\cite{Tian:2008tr}, which indeed would require a reduction in neutrino cooling rates to match the observed surface temperature, as discussed in~\cite{Beznogov:2018fda,Leinson:2019cqv}. However, J173203 is actually a much younger NS, with an age $4 \pm 2$ kyr~\cite{Acero:2015caa,Cui:2016pzm,Maxted:2017xfs}, which is consistent with the standard cooling scenario and does not require unusual pairing gaps. We find that with the corrected NS data the constraint is relaxed. J172054 is a young NS which is naively interesting for axion searches given its well-measured luminosity of $\sim$$(12\pm1) \times 10^{33}$ erg/s~\cite{Ho:2021hwy}. However, this measurement assumes a fixed distance to Earth, and we find that after accounting for the distance measurement and uncertainty~\cite{Potekhin:2020ttj} J172054 is not a powerful probe. J0357, although older than $10^5$ yr, has extremely uncertain luminosity and age measurements, so it is the least constraining NS in our analysis and we do not consider it in the main text. Finally, we include J0538 in the SM because its age is debated and could be a lower value than given in Tab.~\ref{tab:supp-NSs} of $40 \pm 20$ kyr~\cite{Ng:2006vh}. We find that in either case it is not constraining and so exclude it from our main analysis.  {As the ages and luminosities of some of the younger NSs become better understood in the future, it is likely that they will become more sensitive probes of axion-induced cooling.}

\subsection{Effects of nuisance parameters}

\begin{figure}[!htb]
\includegraphics[width = 0.99\textwidth]{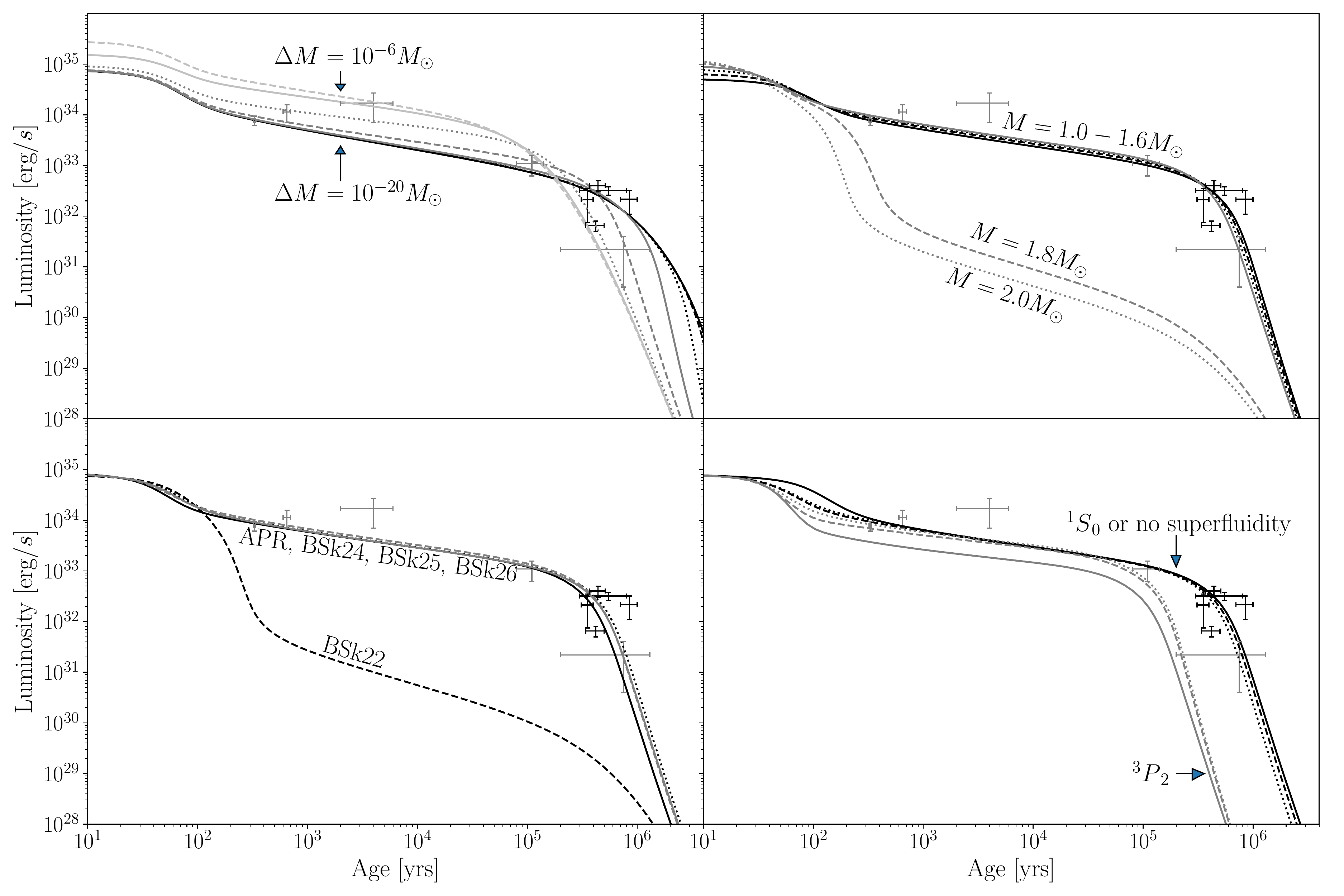}
\caption{As in Fig.~\ref{fig:coolingcurves} except for an example NS with parameters, unless otherwise stated, $M = 1.4 M_\odot$, $\Delta M / M_\odot = 10^{-12}$, the BSk24 EOS, and the \texttt{SFB-0-0} superfluidity model.  Each panel varies the indicated parameter, and no axions are included in the simulations.  On top of the cooling curves we indicated the age and luminosity data for the isolated NSs considered in this work.  The old NSs considered in our fiducial analysis are in black, while the data for the younger NSs that we analyze in the SM are in grey.  }
\label{fig:nuisance}
\end{figure}

In Fig.~\ref{fig:nuisance} we illustrate the effects of the various nuisance parameters that we profile over in our analysis on NS cooling. 
We do not include axion in these simulations, and
unless otherwise stated we fix the NS mass at $M = 1.4 M_\odot$, the mass of light elements in the envelope $\Delta M / M_\odot = 10^{-12}$, the BSk24 EOS, and the \texttt{SFB-0-0} superfluidity model.  On top of the cooling curves which show the surface luminosity as a function of NS age, we indicate the data points for the isolated NSs that we consider in this work. 

In the top-left panel we show the effect of varying $\Delta M$.  Larger values of $\Delta M$ increase the luminosity at early times but can rapidly decrease the luminosity at late times, for ages $\sim$$10^6$ yrs.  The top-right panel indicates illustrates the varying NS mass.  For this particular EOS masses larger than $\sim$1.6 $M_\odot$ reach high enough densities in the core to undergo direct URCA neutrino production and thus cool rapidly.  The bottom-left panel varies the EOS.  Apart from the BSk22 EOS, the the other EOSs give similar results.  The BSk22 EOS is different in that already for $M = 1.4$ $M_\odot$ the direct URCA process is allowed. Lastly, the bottom-right panel shows the effect of the superfluidity model.  The $^1S_0$ and no superfluidity models, which are our fiducial choices, produce similar results.  The $^3P_2$ pairing models that we consider in the SM.  These models undergo rapid cooling at late times when the temperature drops below the $^3P_2$ critical temperature, from PBF neutrino production, and are inconsistent with the M7 data.  We discuss the $^3P_2$ superfluidity models more in the next subsection.

\subsection{$^3P_2$ superfluidity}

In the right panel of Fig. S6 we showed that the $^3P_2$ superfluidity models appear inconsistent with the isolated NS data.  Here, we present further details of our $^3P_2$ superfluidity tests.  We consider four different superfluidity models, which are characterized by their zero-temperature gaps $\Delta$ as functions of Fermi momenta.  (Note that the critical temperatures are proportional to $\Delta$.)  The first three models are denoted as models ``a", ``b", and ``c" from ~\cite{Page:2004fy}, and, in \texttt{NSCool},  as \texttt{SFB-a-T73}, \texttt{SFB-b-T73}, and \texttt{SFB-c-T73}, respectively.  (These \texttt{NSCool} models also self-consistently use the SFB $^1S_0$ neutron gap model and the T73 proton $^1S_0$ gap model.)  However, more recent works indicate that the $^3P_2$ gap may be substantially lower than in these models (see, {\it e.g.}, the recent review~\cite{Sedrakian:2018ydt}).  Of all the recent models reviewed in~\cite{Sedrakian:2018ydt}, the SCGF model with long and short range correlations from~\cite{Ding:2016oxp} has the lowest gap $\Delta$ across all Fermi momenta.  Thus, between the SCGF model and the models ``a", ``b", and ``c" from ~\cite{Page:2004fy}, we span a large range of gaps discussed in the literature for the possible $^3P_2$ pairing, though of course it is also possible that the gap is substantially lower such that $^3P_2$ superfluidity never occurs (as we assume in the main Letter). 

\begin{figure}[!htb]
\includegraphics[width = 0.49\textwidth]{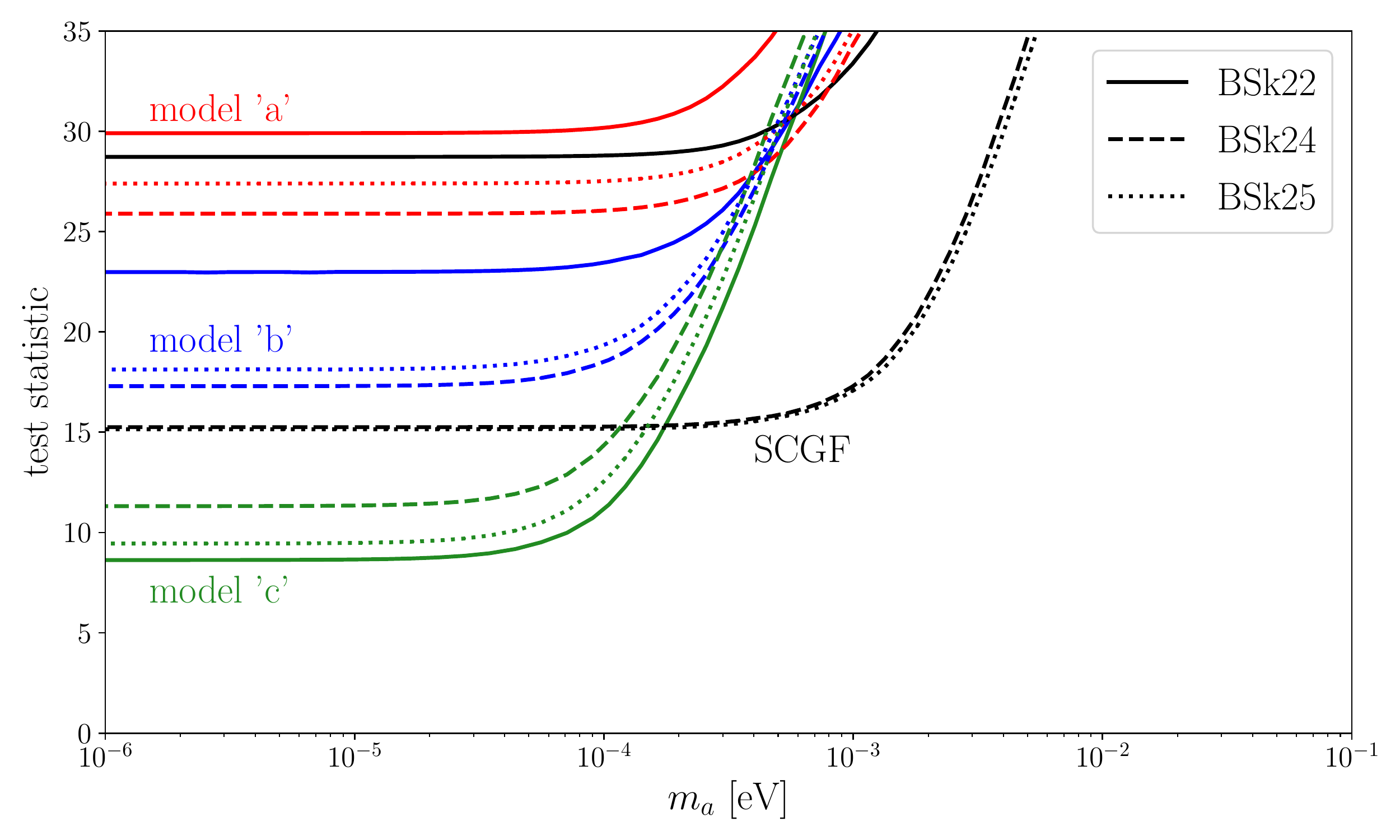}
\includegraphics[width = 0.49\textwidth]{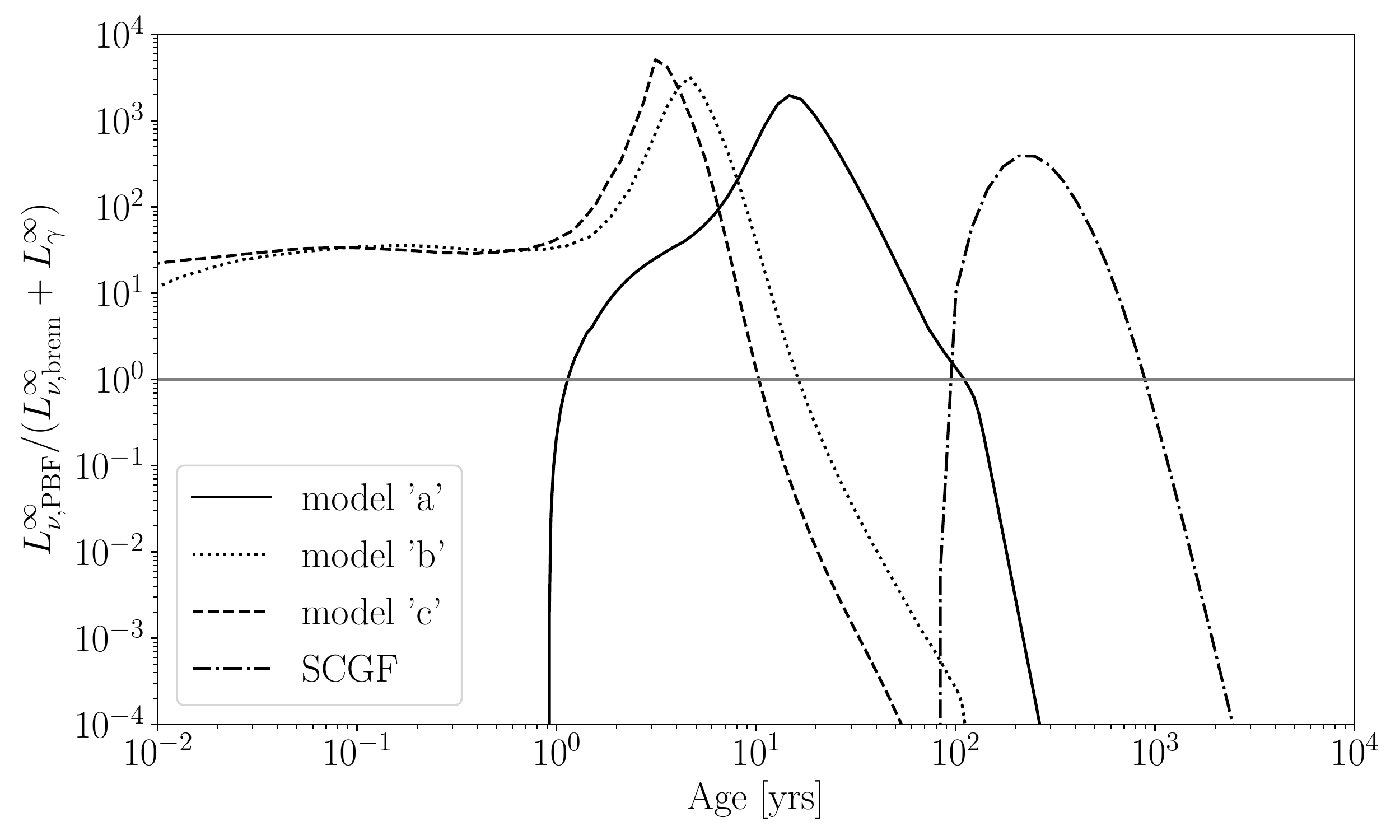}
\caption{ $^3P_2$ superfluidity leads to rapid cooling of the NSs, as illustrated in the right panel of Fig.~\ref{fig:nuisance}.  (Left panel) We show the effects of including axions (for the KSVZ model) with $^3P_2$ superfluidity, for the $^3P_2$ gap models considered in this work.  Note that the SCGF model has the lowest gap and model `c` has the highest gap.  The axion PBF process results in the NSs being much more sensitive to $m_a$, but on the other hand the $^3P_2$ models appear inconsistent with the isolated NS data.  (Right panel) Here, for the case $m_a = 0$, we show the neutrino PBF luminosity relative to the neutrino bremsstrahlung and thermal surface luminosities.  When the neutrino PBF luminosity dominates the NS will undergo rapid cooling. Note that for this figure we fix $M = 1.4$ $M_\odot$.    }
\label{fig:3p2}
\end{figure}

In the left panel of Fig.~\ref{fig:3p2} we show the upper limit test statistics, for different EOS, as a functions of the KSVZ axion mass $m_a$ for the indicated $^3P_2$ gap models.  Note that all of these models are inconsistent with the isolated NS data at more than $\sim$3$\sigma$, as may be inferred from the test statistic at $m_a = 0$ (more precisely, the axion model prefers a negative mass at more than $3$$\sigma$).  On the other hand, these models are much more sensitive to axions due to the PBF axion production mechanism.  In the right panel of Fig.~\ref{fig:3p2} we show how the neutrino PBF process leads to the rapid cooling for the $^3P_2$ models, as seen in {\it e.g.} the right panel of Fig.~\ref{fig:nuisance}. We illustrate the ratio of neutrino PBF luminosity ($L_{\nu, {\rm PBF}}^\infty$) to the bremsstrahlung neutrino luminosity plus $L_\gamma^\infty$.  When this ratio is greater than unity it means that the PBF neutrino process dominates the cooling. Note that since the SCGF model has the lowest gap the $^3P_2$ superfluidity turns on at the lowest temperature of the models considered.  For this example we fix $M = 1.4$ $M_\odot$ and $\Delta M = 10^{-12} M_\odot$.  

\begin{figure}[!htb]
\includegraphics[width = 0.7\textwidth]{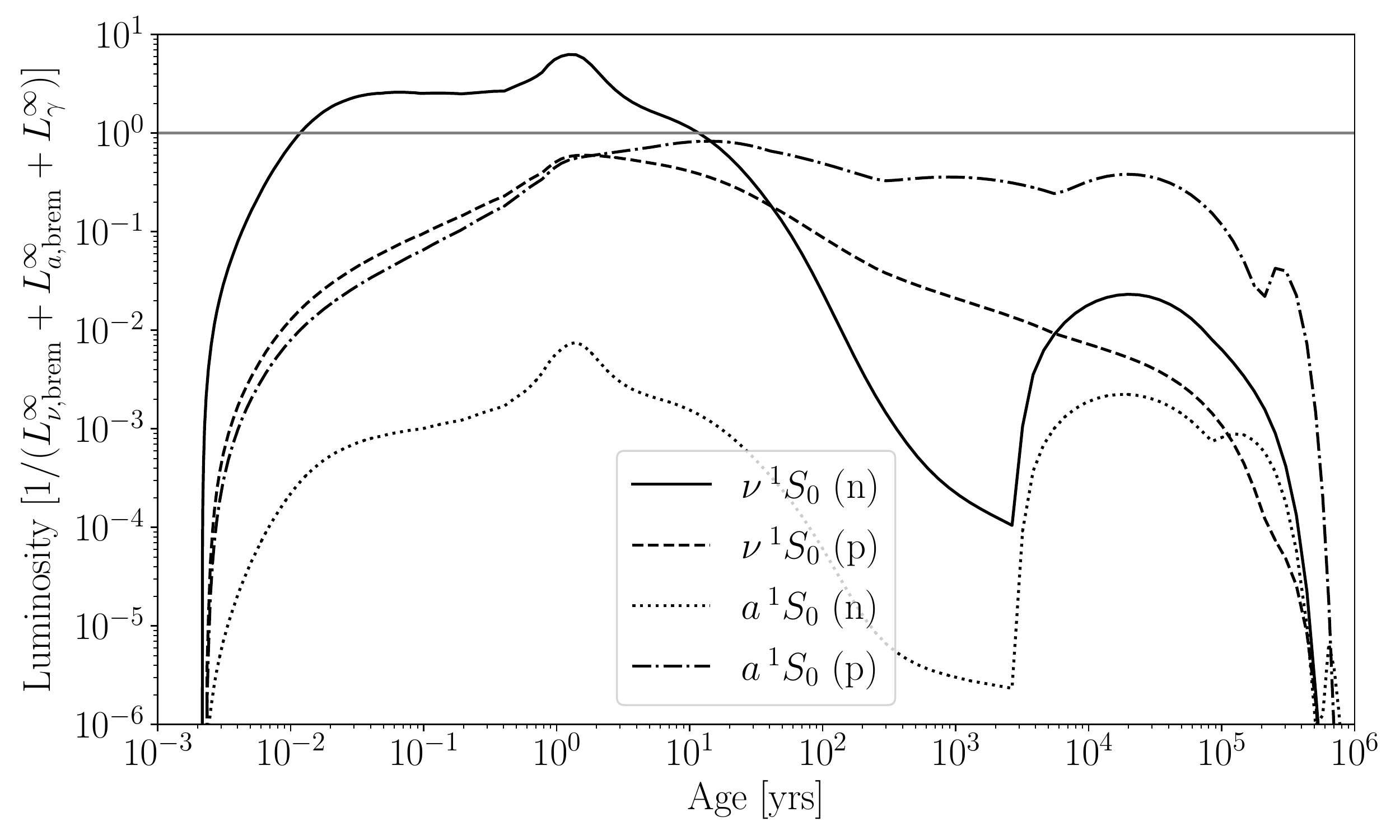}
\caption{
{As in the right panel of Fig.~\ref{fig:3p2} but for the BSk22 EOS, a $1.0$ $M_\odot$ NS, and the superfluidity model with $^1S_0$ neutron and proton pairings but no neutron $^3P_2$ pairing.  We additionally include a KSVZ QCD axion with $m_a = 16$ meV.  We show the ratios of the PBF luminosities for neutrinos and axions produced from neutrons and protons relative to the total bremsstrahlung luminosity plus the thermal surface luminosity. This figure illustrates that the PBF process plays a less important role for the $^1S_0$ pairings than for the $^3P_2$ pairing.}  }
\label{fig:PBF_1s0}
\end{figure}

{
We contrast the right panel of Fig.~\ref{fig:3p2} with Fig.~\ref{fig:PBF_1s0}, where we illustrate similar luminosity ratios for the BSk22 EOS and a $M = 1.0$ $M_\odot$ NS for the model with no $^3P_2$ superfluidity but $^1S_0$ neutron and proton superfluidity, as we consider in our fiducial analyses. (Note that the dependence on the EOS and NS mass is minor, and we also take  $\Delta M = 10^{-12} M_\odot$.)  For that figure we take the KSVZ axion model with $m_a = 16$ meV, at our 95\% upper limit.  We show the ratios of the axion and neutrino PBF luminosities, for neutron and proton production, to the sum of the total neutrino and axion bremsstrahlung luminosities and the thermal surface luminosity.  Comparing this figure with the right panel of Fig.~\ref{fig:3p2} illustrates that the $^3P_2$ pairing, if present, plays a much more important role on the NS thermal evolution than the $^1S_0$ pairings.  
}

\subsection{Sensitivity to the neutron star mass}

\begin{figure}[!htb]
\includegraphics[width = 0.7\textwidth]{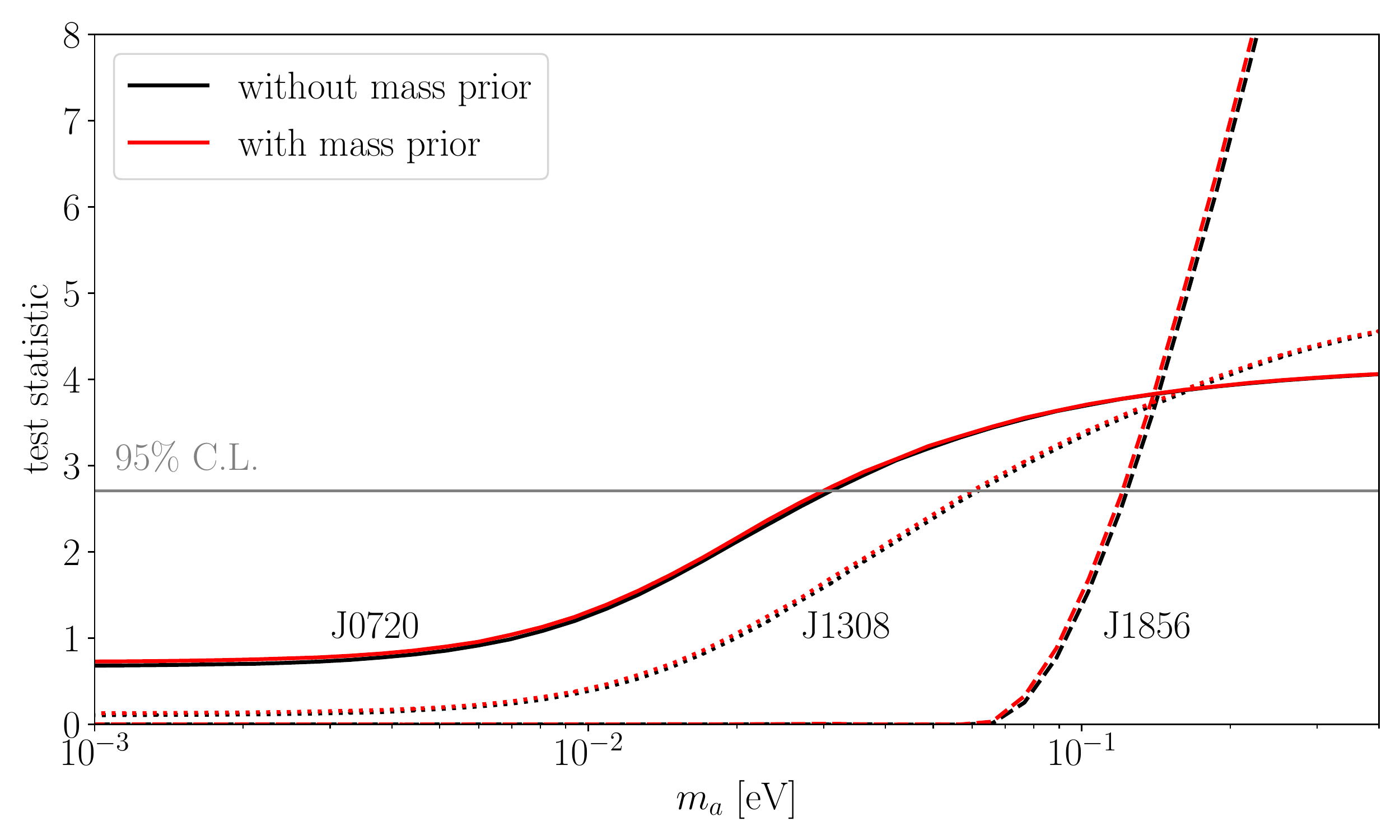}
\caption{Mass estimates are available from~\cite{Tang:2019ign} for the NSs J0720, J1308, and J1856.  Here we consider including these mass estimates as priors in our analysis.  As shown here, however, including the mass estimates makes little difference in our analysis.  }
\label{fig:massprior}
\end{figure}

In Fig.~\ref{fig:nuisance} we show that the NS cooling is not very sensitive to the NS mass unless the mass is large enough for the direct URCA process to become active.  As an additional test of the effect of the NS mass on our final results, we consider including mass priors on J0720, J1308, and J1856.  We do so following~\cite{Tang:2019ign}, who estimate that masses of these NSs as $1.23_{-0.05}^{+0.10}$ $M_\odot$ (J0720), $1.08_{-0.11}^{+0.20}$ $M_\odot$ (J1308), and $1.24_{-0.29}^{+0.29}$ $M_\odot$ (J1856) using gravitational redshift data.  We analyze the cooling data for each of these NSs individually but we add to the likelihood an extra Gaussian contribution for the NS masses, centered on the mass estimates and with standard deviations given by the average of the upper and lower uncertainties from~\cite{Tang:2019ign}.  The results, shown in Fig.~\ref{fig:massprior}, indicate that the upper-limit test statistics are nearly identical with and without the inclusion of the mass priors.

\section{Magnetic field decay}
\label{app:B-field}

In this section consider the possible effects of magnetic field decay on the axion mass upper limits.  We note that despite some dedicated efforts ({\it e.g.},~\cite{1990A&A...229..133H,Miralles:1998na,Page:2000bh,Geppert:2003ir,Arras:2004wq,Cumming:2004mf,Pons:2006ef,Pons:2007vf,Aguilera:2007dy,Popov:2009jn,Vigano:2013lea,Vigano:2021olr}) it is still not standard to include magnetic field decay in NS cooling, in part because the underlying mechanisms that transfer heat from the magnetic field to the NS matter are still not well understood.  For example, while the magnetic field at the surface of a NS may be measured by {\it e.g.} spin-down -- the M7 have field strengths $\sim$$10^{13}$ G -- the magnetic fields within the NSs are not directly observable.  The dissipation of energy from these fields into heat depends not only on the field strength and decay rate but also on the geometry of the fields within the NSs and the dissipative mechanisms, such as the electric conductivity, which are also uncertain.  Still, there is evidence that old ($\sim$Myr-Gyr), isolated NSs, such as the M7, with stronger magnetic fields have higher surface temperatures than those with lower magnetic fields, suggesting that magnetic field decay may play a role in determining the surface temperature of isolated NSs in the epoch after neutrino emission dominates the energy loss, though this is far from conclusive (see, {\it e.g.},~\cite{Pons:2006ef}).  

\begin{figure}[!htb]
\includegraphics[width = 0.7\textwidth]{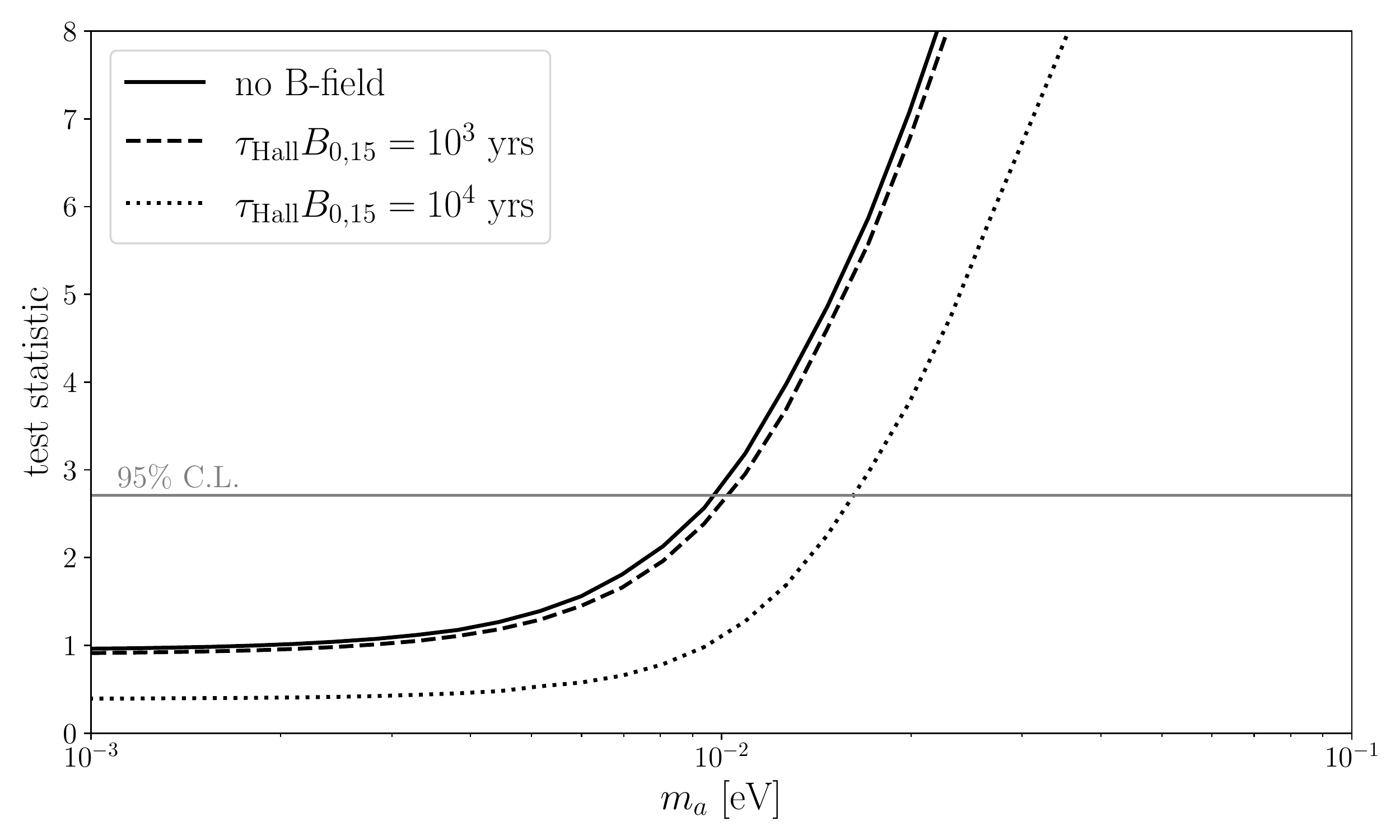}
\caption{We test for the possible effects of magnetic field decay on our axion upper limits by using a parametric magnetic field decay model, where the magnetic field energy is dissipated in the crust as a heat source.  The results depend sensitivity, in this model, on the Hall timescale. Using estimates ({\it e.g.},~\cite{Vigano:2021olr}) that this timescale is less than $\tau_{\rm Hall} =  10^3 \, \, {\rm yr} \,  / B_{0,15}$, with $B_{0,15}$ the value of the magnetic field in units of $10^{15}$ G, we conclude that B-field decay likely plays a subdominant role, as it produces similar results to the no B-field scenario.  On the other hand, an increased Hall timescale, as illustrated, could weaken the upper limits.  In these analyses we allow for initial magnetic fields up to $10^{16}$ G that fill the entire crust. }
\label{fig:Bfield}
\end{figure} 

While complex 2D simulations of the coupled magneto-thermal evolution are now available (see~\cite{Vigano:2021olr} for the state-of-the-art), we take a simpler approach in this work since, as we will show, the effect of magnetic field decay appears to be subdominant compared to other effects that we consider (though this should be checked with dedicated magneto-thermal simulations incorporating axions).  We follow instead the effective approach from {\it e.g.}~\cite{1990A&A...229..133H,Pons:2006ef} and assume that the heat $H$ from magnetic field decay (see~\eqref{eq:cooling} for the relation of $H$ to $L_\gamma^\infty$) is given simply by $H = -V_{\rm eff} B \dot B$, where $V_{\rm eff}$ is the effective volume of the NS that supports the magnetic field $B$, with time derivative $\dot B$.  As is standard, we assume that the dominant Joule heating arises from Ohmic dissipation in the crustal-confined magnetic fields, such that in our modified version of {\rm NSCool} we only include energy injection from $H$ into the crust.  The magnetic energy is likely dissipated in hot-spots in the crust~\cite{Pons:2006ef}, such that $V_{\rm eff}$ is much smaller than the full volume of the crust.  The double blackbody fits for the M7 indicate that the hot-spots have size of a few km across~\cite{Beznogov:2021ijc,Ho:2006uk,2012AA...541A..66S,2011AA...534A..74H,Pires:2019qsk}.  In these hot spots the magnetic field could be $\sim$10 times higher than the poloidal values~\cite{Pons:2006ef}.    To be conservative we assume $V_{\rm eff}$ is the full crustal volume, which is typically around $1/3$ of the total NS volume, and we consider initial magnetic fields up to $10^{16}$ G that fill the entire crust.

Following~\cite{Aguilera:2007dy} we use the phenomenological parameterization of the magnetic field decay
\es{eq:dBdt}{
B(t) = B_0 {e^{-t / \tau_{\rm Ohm}} \over 1 + (\tau_{\rm Ohm} / \tau_{\rm Hall})(1 - e^{-t / \tau_{\rm Ohm}} )} \,
}
with $B_0$ the initial magnetic field strength, $\tau_{\rm Ohm}$ the Ohmic dissipation timescale and $\tau_{\rm Hall}$ the Hall diffusion timescale.
Ref.~\cite{Aguilera:2007dy} suggests $\tau_{\rm Ohm} \sim 10^6$ yr and $\tau_{\rm Hall}$ up to $\tau_{\rm Hall} \sim 10^3 \, \, {\rm yr} \, / B_{0,15}$, with $B_{0,15} \equiv B_0 / 10^{15} \, \, {\rm G}$, though the Hall timescale could be faster.  As a test we fix $\tau_{\rm Ohm} = 10^6$ yr (this timescale plays a less important role than the Hall timescale) and $\tau_{\rm Hall} =  10^3 \, \, {\rm yr} \,  / B_{0,15}$.  In Fig.~\ref{fig:Bfield} we show the result of adding this heating contribution in the context of the KSVZ axion model upper limit test statistic profile.  We compare the new result to that with no B-field decay, as considered in the main Letter.  The difference between the two results is minimal.  Smaller values of $\tau_{\rm Hall}$ lead to smaller differences, indicating that magnetic field decay likely plays a subdominant role.  As a test, however, we consider increasing $\tau_{\rm Hall}$ by an order of magnitude to $\tau_{\rm Hall} = 10^4 \, \, {\rm yr} \, \, / B_{0,15}$.  In Fig.~\ref{fig:Bfield} we show that in this case the upper limit may be weakened by $\sim$50\%.  Given the potential relevance of magnetic field decay to the axion upper limits, this process should be investigated further in future works using more sophisticated treatments of the B-field decay.

\end{document}